# Toward a More Natural Expression of Quantum Logic with Boolean Fractions[*]

Philip G. Calabrese
calabres@spawar.navy.mil

Joint & National Systems Division (Code 2737)
Space & Naval Warfare Systems Center (SSC-SD)
San Diego, CA 92152-5001

**Abstract**: The fundamental algebraic concepts of quantum mechanics, as expressed by many authors, are reviewed and translated into the framework of the relatively new non-distributive system of Boolean fractions (also called conditional events or conditional propositions). This system of ordered pairs (A|B) of events A, B, can express all of the non-Boolean aspects of quantum logic without having to resort to a more abstract formulation like Hilbert space. Such notions as orthogonality, superposition, simultaneous verifiability, compatibility, orthoalgebras, orthocomplementation, modularity, and the Sasaki projection mapping are translated into this conditional event framework and their forms exhibited. These concepts turn out to be quite adequately expressed in this near-Boolean framework thereby allowing more natural, intuitive interpretations of quantum phenomena. Results include showing that two conditional propositions are simultaneously verifiable just in case the truth of one implies the applicability of the other. Another theorem shows that two conditional propositions (a|b) and (c|d) reside in a common Boolean sub-algebra of the non-distributive system of conditional propositions just in case b=d, that their conditions are equivalent. Some concepts equivalent in standard formulations of quantum logic are distinguishable in the conditional event algebra, indicating the greater richness of expression possible with Boolean fractions. Logical operations and deductions in the linear subspace logic of quantum mechanics are compared with their counterparts in the conditional event realm. Disjunctions and implications in the quantum realm seem to correspond in the domain of Boolean fractions to previously identified implications with respect to various naturally arising deductive relations.

---

[*] Support for this work under the SSC-SD In-house Independent Research Program is gratefully acknowledged.

# Table of Contents

# List of Figures

# List of Tables

# 1. Introduction & Background

Among the most difficult things to accept about quantum mechanics is its need to discard 2-valued Boolean logic in order to adequately describe quantum situations and events. Instead of the old familiar Boolean algebra of subsets under union, intersection and complement, quantum mechanics requires a different logic. One reason is that quantum observation alters the object observed. Thus quantum observables are operators, not merely passive variables being measured. Also, experimental conditions for the measurement of one variable may preclude the conditions necessary for the measurement of another variable. Thus simultaneous measurements are sometimes impossible. This never happens in Boolean logic. Hence there is need for an alternative logic in quantum mechanics. H. Putnam [Put76, p51] in describing the physical propositions of quantum mechanics says, " … the corresponding lattice of physical propositions is not 'Boolean'; in particular, distributive laws fail."

(The main purpose of this paper is to promote the relatively new, *non-distributive* system of Boolean fractions, conditional events, as a more natural algebraic context in which to express quantum mechanics. Comments in support of this contention are interspersed throughout this introduction as the writings and results of various quantum researchers are reviewed.)

**1.1 Quantum Foundations.** So when J. von Neumann [Neu32] was searching for a way to formalize the quantum mechanics initiated by Planck [Pla00]], Einstein [Ein06], de Broglie [Deb23], and Schrödinger [Sch26] he could not use Boolean algebra to represent quantum events or propositions because Boolean algebra is wholly distributive while quantum events display non-distributive logical relationships. The non-commutativity of some quantum observations (measurements) implies the non-distributivity of quantum operations and therefore precludes exclusive use of Boolean algebra to describe quantum physical situations and interactions.

Together with G. Birkhoff [Bir36] von Neumann refined and formulated what is now known as "orthodox" quantum logic – a complete, normed, vector space, also an orthocomplemented lattice. This and equivalent formulations [Sch26], [Dir26] have been quite successful at calculating the probability that a variable of a quantum system in a given state $\psi_1$ will take a specified value and the state $\psi_1$ will change to another state $\psi_2$ as the variable is measured. The motivations for these alternative formulations have usually included a desire to make quantum mechanics somewhat more natural, more easily interpreted in terms of recognizable physical concepts.

This effort has been complicated by the many strange features that apparently must be incorporated into any account of quantum phenomena. Among these are the trajectory interference phenomena associated with the passage of even a single particle through either of two narrow slits [Fey65], and the so-called non-local reality effects railed against by A. Einstein et al [Ein35] such as the violation of Bell's inequality [Bel64] and more general entanglement effects [Pit89].

B. Coecke, D. Moore and A. Wilce [Coe01] provided a good overview of "operational quantum logic". They started by describing the approach of J. von Neumann and the logic of linear projections noting that while not wholly Boolean, "the sub-ortholattice generated by any commuting family of projections is a Boolean algebra." These non-commuting projections arise from the non-distributive operations, as first pointed out by von Neumann [Bir36, p839]. Notice that any set of commuting projections (variables) reside in a sub-algebra that is Boolean.

In discussing Mackey's account [Mac57] of quantum mechanics as a probability calculus, Coecke, et al [Coe01], pointed out how "once one entertains the idea that the testable propositions associated with a physical system need not form a Boolean algebra, the door is opened to a huge range of other possibilities. It then becomes a matter of urgency to understand *why* nature (or we) should choose to model physical systems in terms of projection lattices of Hilbert spaces, rather than anything more general." One can also ask why not something simpler and less abstract? One reason is that the non-distributive system of Boolean fractions did not exist when quantum mechanics was being formulated.

In an amusing paper Fuchs [Fuc02, p1] asked, "… when will we ever stop burdening the taxpayer with conferences devoted to the quantum foundations?" His answer: Not until a "means is found to reduce quantum theory to two or three statements of crisp physical (rather than abstract, axiomatic) significance." He listed several quantum "religions with their priests pitted in holy war" against each other. Fuchs suggested that like relativity theory, quantum theory needs someone like Einstein to boil down the abstract mathematics of quantum mechanics into simple physical statements that account for the strange quantum formulas just like Einstein's two relativity axioms imply the otherwise mysterious, empirically discovered Lorentz transformations. However Fuchs admitted to asking for much more than he provided as he affirmed his faith in the efficacy of the positive operator-valued measures (POVM) approach.

**1.2 Probabilistic Metric Spaces.** One theoretical off-shoot of quantum mechanics, an idea introduced by K. Menger [Men42], [Men51] was probabilistic metric spaces [Cal68], [Cal79] - spaces of geometric points whose distances are postulated to take on values with certain probabilities. Perhaps the strange particle-wave affects of quantum mechanics might be more intuitively grasped by postulating a geometry whose points have probabilistic distance expressed in terms of cumulative distribution functions for each pair of points. Khrennikov [KhrDec01, p2] related similar ideas back to N. Bohr's "statistical model of physical reality." However, the development of this idea did not shed much light on quantum mechanics because Kolmogorov's conditions for a joint distribution were assumed for any subset of distribution functions, thereby making any two distance-variables expressible as random variables on a common sample space.

**1.3 Inadequacies of Probability Theory.** In the meantime, it became clear [Cal75], [Lew76], that probability theory was incomplete in not having an adequate ordered pair (a|b) that could carry the conditional probability P(a|b) and also operate with the other logical operations of "and", "or" and "not". In particular the usual translation of "if b then a" as "either a or not b", the so-called material conditional, is quite inappropriate as soon as either the premise "b" or the statement itself is uncertain: $P(a|b) < P(a \text{ or not } b)$ except when $P(b) = 1$ or $P(a|b) = 1$.

This anomaly is the reason why in standard probability theory and practice the underlying logical relationships between events or propositions are relegated to positions inside the probability function. For example, instead of asserting "b" and also "if b then a" and then passing to "a" by applying a probabilistic form of modus ponens, in probability theory one writes $P(a) = P(b)P(a|b)$ and never writes any logical forms like "a given b" by themselves.

Lewis [Lew76] proved that only in trivial Boolean algebras could (a|b) be assigned P(a|b) and also be an event in the original Boolean algebra containing the events a and b. However since the

appearance of [Cal87] there has been available an extension of Boolean algebra to Boolean fractions (a|b) and these fractions can be assigned P(a|b). Subsequently this theory was extended to deduction with uncertain conditionals [Cal91], [Cal02] & [Cal03A].

Boolean fractions are order pairs of propositions (a|b) representing "a given b" or "a if b" that can be combined with the three usual Boolean logical operations and even allow iterations, fractions of fractions, such as "if c then (if b then a)", "if (b given c) then a", and most generally, "if (c given d) then (a given b)".

These Boolean fractions do for propositions & events what numerical fractions do for integer arithmetic. They dramatically expand what can be accurately represented. Imagine trying to represent and compute the distances on a continuous number line using only whole number lengths. Something quite akin to this is routinely being done when uncertain, conditional information is represented by Boolean propositions or events.

It is safe to say that were Boolean fractions available when Birkhoff and von Neumann were formulating quantum logic they would have tried to use them because the algebra of Boolean fractions is non-distributive.

**1.4 Boolean Fractions and Conditional Events.** Perhaps the first person to seriously consider defining operations on ordered pairs of Boolean propositions for purposes of applying them to quantum mechanics was G. Schay [Sha68]. Although philosopher E. Adams [Ada66] had already defined his so-called "quasi operations" on conditional statements, his development was barely algebraic. For instance he called his "and" and "or" operations "quasi" merely because they were not monotonic. That is, a conjunction did not always imply its components, and a disjunction was not always implied by its components. Because of these properties Adams said they were not true conjunction or disjunction operations. Nor was Adams trying to apply his conditionals to quantum mechanics.

B. Definetti [Def36], who was first to state the three-valued nature of conditionals, pursued the problem of generating probability distributions with partial information, but he did not define operations on his conditionals. One other early effort along these lines was that of Z. Domotor [Dom69], whose Ph.D. dissertation under P. Suppes considered ordered pairs of propositions for qualitative probability relations.

**1.5 Compatibility, Distributivity and Simultaneous Verifiability.** As expressed by Putnam [Put76, p51], "The whole function of the linear spaces used in quantum mechanics is to provide a convenient mathematical representation of the lattice of physical propositions….". Concerning simultaneously verifiable or compatible propositions several equivalent ways have been discovered to express their relationship:

In a complete orthocomplemented weakly modular lattice Piron [Pir76, 25] defined two propositions b and c to be *compatible* if the sub-lattice generated by them and their negations is distributive.

Jauch [Jau76, p133] defined two lattice elements a and b to be *compatible* if the smallest sub-lattice which contains a and b and their negations is Boolean.

A. Fine [FinA73] proved that two quantum operators commute just in case they have a joint distribution. Later he proved [FinA82] that Bell's inequalities are sufficient for the existence of a joint probability distribution between the measurement variables. Also see [Sup96, p120].

In each of these definitions or theorems compatible propositions act like Boolean propositions. As will be shown, the system of Boolean fractions is well suited to express these relationships.

Piron [Pir76, p22] offered a simple example illustrating non-distributivity as related to the occurrence of photons passing through polarizers of different orientation. While light will pass one polarizer, it will not pass two unless they are oriented 180 degrees to one another. Somewhat more clearly, Gudder & Nagy [Gud01, p1125] stated how one ordering of 3 polarizers eliminates all light but that interference allows a different ordering to pass some light.

**1.6 States, Pure States and Indicator functions.** [Pir76, p36] said that a "state" in a quantum system is represented by an "atom" in the propositional system.

Coecke, et al, [Coe01, p9] quote Jauch and Piron [Jau69, section 5] saying that states are maximal sets of actual properties of the system, and also (p14) that "states correspond exactly to atoms" of the lattice.

So in the Bohmian, so-called "hidden variable formulation" [Boh52], a state seems to be equivalent to a complete assignment of a value to each quantum variable. The collection of all such complete assignments corresponds to the "situation space", a model familiar from various formulations of uncertainty reasoning for military and other complicated situations of variables and their values. By attaching a probability to each such complete assignment, so that the sum of all of them is 1, a unit vector is thereby defined whose components are the complete assignments. The set of all possible ways to attach probabilities to this set of complete assignments is the collection of all possible unit vectors each of whose components is a complete assignment of values to variables. So the set of all states can be considered to be the set of all unit vectors whose components are each a complete assignment of a value to each variable. The only question is whether such simultaneous assignments have meaning, or are really meaningless as the so-called Copenhagen interpretation holds.

Jauch [Jau76, p129] said, "Every state of the system defines a new kind of measure." He defined a pure state as a measurable indicator function $\delta_\omega$ defined on the measurable subsets A of a sample space $\Omega$ as:

$$\delta_\omega(A) = \begin{cases} 1 & \text{for } \omega \in A \\ 0 & \text{for } \omega \notin A. \end{cases}$$

Here, $\Omega$ can be the collection of complete assignments and $\omega$ one of them represented by the indicator function $\delta_\omega$.

**1.7 Context, Conditions, and Experimental Arrangements.** L.E. Ballentine [Bal85, p885] made some very perceptive remarks about changing experimental conditions and the interpretation to be given to the inability in quantum mechanics to make some simultaneous measurements: "Since different apparatuses are used to measure position than to measure momentum, one will be dealing with $P(A|C_q)$ and $P(B|C_p)$, where $C_q$ includes the configuration of any measuring apparatus of the position measuring device and $C_q$ includes the configuration of the momentum measuring device."

This perception followed D. Bohm [Boh80, p74]: "The above discussion of the meaning of a measurement then leads directly to an interpretation of the indeterminacy relationships of Heisenberg. As a simple analysis shows, the impossibility of theoretically defining two non-commuting observables by a single wave function is matched exactly, and in full detail, by the impossibility of the operation together of two overall set-ups that would permit the simultaneous experimental determination of these two variables. This suggests that the non-commutativity of two operators is to be interpreted as a mathematical representation of the incompatibility of the arrangements of apparatuses needed to define the corresponding quantities experimentally."

This has been echoed in different contexts such as by P. Suppes [Sup90, p293]: "If we avoid noncommuting variables in quantum mechanics, then probability is classical." And [Sup90, p296] "My central thesis … is that quantum mechanics is consistent with various stochastic extensions if we ignore computations on non-commuting variables." Suppes has an averaging interpretation in mind [Sup74] but there is another way to proceed, namely, with Boolean fractions, which are generally non-distributive and yet reduce to classical Boolean algebra when conditions are fixed.

Similarly, J. Hilgevoord [Hil01, p14] expressing Bohr's argument about measuring a momentum by measuring a recoil, has said, "Since a measuring instrument cannot be rigidly fixed to the spatial reference frame and, at the same time, be movable relative to it, the experiments which serve to precisely determine the position and momentum of an object are mutually exclusive."

Hardegree [Har76, p55] discussed objections in the literature to quantum logic being considered a true logic as it lacks a "conditional operation by means of which the modus ponens deduction scheme can be incorporated into quantum logic". Well, the non-distributive algebra of Boolean fractions does in fact have such a conditional operation and it also supports deduction using modus ponens [Cal87, p228].

Domotor [Dom76, p176] used something called "Boolean atlases" to cover quantum lattices with simple Boolean lattices in an attempt to provide easier interpretations of quantum theories. Again the idea here seems to be to provide changing contexts. But that is precisely what Boolean fractions do.

Aerts et al [Aer00B] have produced macroscopic examples that violate Bell's inequality by identifying events that are not really the same. They ask, "Is the violation in the microscopic world perhaps due to a lack of distinguishing events that are in fact not identical?" Whether or not one accepts this suggestion for the resolution of Bell's inequality and non-locality in quantum mechanics, distinguishing quantum events with different conditions is easily accomplished in the system of conditional events.

In discussing the role of experimental conditions, T. Fine [FinT76, p185] quoted A. Messiah's 1962 text *Quantum Mechanics* (p154): "As a consequence, evidence obtained under *different experimental conditions cannot be comprehended within a single picture.*" Fine went on to say "When … the operators representing observables do not commute then the order of performance of the 'joint' measurement affects the outcome, and thus there is said to be no joint observation."

In a flurry of papers [KhrJun01], [KhrDec01], [KhrFeb02] and [KhrNov02], A. Khrennikov has made very pertinent remarks about the role of changing *context* in quantum mechanics as related to different experimental arrangements. In [KhrFeb02, p7], [KhrNov02, p3] he claimed to show that contrary to common opinion, EPR-Bohm type "correlations can be obtained in the *local realistic approach* if we carefully combine probabilities corresponding to different physical contexts." Again, whether or not one agrees with Khrennikov's conclusion that classical "local reality" can be maintained in quantum mechanics by carefully treating different physical conditions, clearly the role of changing physical contexts is a crucial aspect of quantum mechanics that should be more carefully addressed before more radical methods of reconciling quantum EPR-type experiments with logic are employed.

An emphasis on conditions is also evident in R. D. Sorkin's use of different "histories" [Sor97, p8]: "Having decided to interpret probability in terms of preclusion, one still has the further task of incorporating the fact that what is precluded is not fixed once and for all, but rather 'changes with time'; or in other words the fact that the future is *conditioned* on the past. … To do this one needs to be able to make conditional statements of the form, "If the past has such and such properties then such and such a future possibility is precluded."

However in [Sor94, p4] he seemed to accept the questionable notion in the two-slit experiment that the condition of having two slits open is the disjoint union of the two conditions - one slit open and the other slit closed, or vice versa. However as was pointed out by Ballentine [Bal86] and others, having two slits open is evidently not the same as the union of the two conditions of having just one or the other slit open. For instance, if some unknown influence such as De Broglie-Bohm "pilot waves" change the spatial environment by mutual interference when two slits are open, versus when just one is open, then the trajectory of a particle going through either slit would be different from the union of the trajectories when just one slit is open. De Broglie [Deb60, pp89-90] early promoted these ideas as recounted in his very clear 1960 book.

**1.8 Contextuality**. Khrennikov explained [KhrNov02, p3] that one of the main reasons that "contextualism" was not successful in deriving quantum statistics is that "the standard probabilistic formalism based on Kolmogorov's axiomatics [Kol56] was a fixed context formalism. This conventional probabilistic formalism does not provide rules of operating with probabilities calculated for different contexts. However in quantum theory we have to operate with statistical data obtained for different complexes of physical conditionals, contexts"

Now this is exactly what the (non-distributive) algebra of Boolean fractions provides [Cal87, Cal91] - an extension of Kolmogorov's probability theory to propositions or events with variable conditions. Subsequently a theory of deduction for uncertain conditionals [Cal02] has been provided for these Boolean fractions, also called conditional events. Recently, these methods for operating on indicator functions with different domains were extended to arbitrary real-valued functions including conditional random variables [Cal03].

S. Goldstein [Gol96A, p11] quoted J. S. Bell [Bel66, p166] on the topic of contextuality saying that the "misuse of the word 'measurements' makes it easy to forget" that the results of quantum measurements "have to be regarded as the joint product of 'system' and 'apparatus', the complete experimental set-up". Bell said that forgetting this has led people to expect that "the 'results of measurements' should obey some simple logic in which the apparatus is not mentioned. The resulting difficulties soon show that any such logic is not ordinary logic. It is my impression that the whole vast subject of 'Quantum Logic' has arisen in this way from the misuse of a word."

Given these observations from Bell and Goldstein, who have deeply examined these issues, it seems clear that the thesis presented here that Boolean fractions, an algebra of propositions each with its own condition, can provide a near Boolean framework for the representation of quantum mechanics. Compared to the standard abstract Hilbert space quantum logic with its standard Copenhagen interpretation that more or less asks scientists to check their physical intuitions at the laboratory door, the prospect of a more natural formulation has great attraction and should be thoroughly explored. It might even save the taxpayers a lot of money [Fuc02, p1] by reducing the need for so many conferences on quantum foundations!

In a somewhat contrary direction D. Gillespie [Gil86, p889] disproved the possibility of describing quantum mechanics in terms of "a statistical ensemble interpretation" wherein "the quantum mechanical state vector is assumed to define a statistical ensemble of identically prepared systems, each of which has precise values for all its observable variables, and the act of measurement is equivalenced to a straightforward sampling of that ensemble." Gillespie offered a counter-example that appears to be impossible to represent in this way thus lending support to the Copenhagen interpretation of quantum mechanics.

However, the example and proof he offered is strongly dependent upon the summation formula for disjoint events: For disjoint propositions (or events) A and B with the same condition C, we always have $P((A|C) \vee (B|C)) = P(A|C) + P(B|C)$. However this formula does not in general hold when A has one condition $C_1$ and B has a different condition $C_2$. In this situation, $P((A|C_1) \vee (B|C_2))$ is not in general equal to $P(A|C_1) + P(B|C_2)$. For instance, the probability of getting "1 given an odd roll of a die or 2 given an even roll" is 2/6 not 1/3 + 1/3, even though "1" and "2" are disjoint as are "even" and "odd". Also see Theorem 2.13.

**1.9 Quantum Logical Operations.** In a clear exposition Aerts et al [Aer00A] have explained why in the standard formulation the quantum disjunction and negation operations are not classical. Using both the subspace representation and the projection representation, they also explained how the quantum deduction and conjunction operations behave classically. They recounted macroscopic examples of quantum phenomena [Aer97] produced by having "non-local" interactions between parts of the system. This section was introduced with the statement (p3) that "the reason the quantum logical disjunction does not behave classically is that, for two propositions a, b, the quantum entity can be in a state p, such that $a \vee b$ is true without a being true or b being true". However, does not something very similar happen when in classical probability $P(a \vee b)$ equals 1 without either a or b having probability 1? On the other hand this cannot happen when the state (model) ω is atomic since then a ∨ b is true only if either a is true in ω or b is true in ω.

**1.10 Non-Contextuality**. While many researchers have seen careful attention to context as one key to a more natural expression of quantum mechanics, others have emphasized the strong "non-contextuality" aspect of the standard quantum mechanical formulation.

For instance, working in the context of incompatible gambles, Pitowski [Pit02, p5] has made it clear that in the standard quantum mechanical formulation, if $\mathcal{B}_1$ and $\mathcal{B}_2$ are two Boolean algebras representing two incompatible measurement contexts having a common event F, then $P(F|\mathcal{B}_1) = P(F|\mathcal{B}_2)$. He has pointed out that this is quite different from the usual assignments of conditional probabilities, which in general would have $P(F|\mathcal{B}_1) \neq P(F|\mathcal{B}_2)$. He has also stressed that one cannot assume that $\mathcal{B}_1$ and $\mathcal{B}_2$ are sub-algebras of a larger Boolean algebra $\mathcal{B}$ containing F because there are "sufficiently rich games which cannot be imbedded in a Boolean algebra without destroying the identities of the events and the logical relations between them."

Fuchs [Fuc02, p14] explained that A. Gleason's celebrated theorem [Gle57] shows that the standard quantum mechanical probability rule is "the only rule that satisfies a very simple kind of non-contextuality for measurement outcomes." Non-contextuality, he explained [Fuc02, p20], is simply the identification of outcomes when their probabilities are equal regardless of the initial beliefs about the system or the differences in the design of the apparatuses used to determine the truth of the outcome.

According to Goldstein [Gol01, p17] the Kochen-Specker theorem [Koc67], Gleason's earlier theorem [Gle57], Bell's theorem [Bel64], and several other "no hidden variable" results all "show that any hidden-variables formulation of quantum mechanics must be *contextual*." It must violate the non-contextuality assumption, as Bell put it in 1987 [Bel87, p9], "that measurement of an observable must yield the same value independently of what other measurements may be made simultaneously."

Thus, non-contextuality is a stringent condition that narrows the possible representations of quantum events to that of the standard formulation. However non-contextuality may very well be a peculiarity of the Hilbert space model of quantum phenomena rather than a fundamental physical law that must be incorporated into any representation of quantum mechanics. Considering that quantum physical experimental observations are usually highly statistical in nature, the non-contextuality of the orthodox formulation might be an artifact of the Hilbert space representation rather than something inherent in the physics.

**1.11 Hidden Variables and Non-Contextuality**. Pitowski [Pit02, p16] observed that in Bohm's theory "the observable $S_x^2$ in the x,y,z context is not really the same as the $S_x^2$ in the x',y',z context. Nevertheless, the Bohmians consider $S_x^2$ as one single *statistical* observable across contexts." The reason, he said, is that the *average* outcome of $S_x^2$, over different initial positions with the appropriate density $|\psi(x,0)|^2$ given by the pilot wave, is in fact context independent. Although, as Pitowski said, this sometimes requires a non-local reality model in the Bohmian representation, there seems to be no insurmountable reason that the Bohmian position about contextuality cannot be adopted.

Aerts et al [Aer97, p794] made some careful distinctions about the kinds of hidden variable theories that are possible in view of various proofs of the supposed impossibility of such a theory. They stated that there is one central assumption of all of these "no-go" theorems: "The hidden variables have to be hidden variables of the *state* of the physical entity under consideration and specify a deeper underlying reality of the physical entity itself, independent of the specific measurement that is performed." But they also allowed "there exists always the mathematical possibility to construct so-called hidden variables to depend on the measurement under consideration (e.g. the spin model proposed by Bell [Bel66])". Thus a contextual hidden variable model is still possible.

D. Durr et al [Dur95, p10] have embraced hidden variables and non-local effects. Concerning the instantaneous pilot or guiding wave function of Bohmian mechanics, they wrote, "We propose that the wave function belongs to an altogether different category of existence than that of substantive physical entities, and that its existence is nomological rather than material."

Allowing for the possibility of missed detections, so that "variables can take on a third value, 'D', corresponding to an inherent 'no show' or defectiveness", L. E. Szabó and A. Fine [Sza02] exhibited examples they said are local hidden variable realizations of the Greenberger-Horne-Zeilinger (GHZ) experiment. They used the inequalities developed by Barros & Suppes [Bar00] for just these detector inefficiencies and for GHZ experiments. Here again, the notion of a 3-valued logic, or equivalently, the notion of conditional events (Boolean fractions), seems to naturally find its way into quantum mechanics.

With so many conflicting contentions in the literature about the possibility of a hidden variable formulation of quantum mechanics, it is hard to feel confident about the arguments from any one of the competing camps. While I am sympathetic with the non-local, hidden variable formulation championed by S. Goldstein, the use of Boolean fractions should help to resolve the question of whether non-locality can be eliminated by a more careful labeling of conditions under which certain variables are measured.

**1.12 Local Reality**. It seems rather strange to me that the notion of "non-locality", the influence of distant objects at faster than light speed, is so severely criticized and shunned. The famous Isaac Newton was uncomfortable with the notion. On the other hand Newton's own theory of gravitational attraction is non-local, even though that non-local influence falls off inversely with the square of the distance. Nevertheless, according to T. Van Flandern [Fla98, p4] it has been known for a long time that there can be little or no *delay* of gravitational attraction between, say, the sun and earth. Were the earth-sun gravitational attraction to experience a speed-of-light delay, it would form an angular momentum increasing, force couple that would double the radius of earth's orbit about the sun in 1200 years! Van Flandern [Fla98, p4] estimated the present lower bound for the speed of propagation of gravity to be $2 \times 10^{10}c$, where c is the speed of light.

As quoted by Van Flandern [Fla98, p1] from [Hof83], Newton expressed it this way: "That one body may act upon another at a distance through a vacuum, without the mediation of any thing else, by and through which their action and force may be conveyed from one to the other, is to me so great an absurdity, that I believe no man who has in philosophical matters a competent faculty of thinking, can ever fall into it." Of course, Newton objected to one body acting on another "without the mediation of any thing else". But those who advocate action at a distance do postulate "something else" as mediator, albeit an instantaneous something. So perhaps Newton would accept that competent thinkers could hold such a concept.

A. Einstein thought the idea of action at a distance "spooky" but his own theory of gravitation in which the nearby space is curved by a local mass is still non-local in the sense that the bending-of-space influence must also be essentially instantaneous. It does not matter that in one model the influence is taken to be a force attraction and in the other model taken to be a geometric space-bending influence. The influence is still considered to be instantaneous across great cosmic distances. Saying that it is geometric doesn't make it local unless all geometry is local. Einstein also had another reason for disliking action at a distance; it tended to call into question his claims that simultaneity is completely relative to the observers of events. For if there are instantaneous actions at a distance, then there is a sense in which simultaneity has meaning.

According to Fuchs [Fuc02, p10], the "rigid connection of all nature" while embraced and even glorified by the Bohmians, would make the notion of separate systems meaningless, and so therefore it must be rejected. Now this is obviously way too strong. If the connectedness of "separated systems" could be quantified, for example, by the instantaneous "pilot waves" of the Bohmian model, then absolute separateness would just have to be qualified. Nowadays with the "big bang" cosmology embracing action-at-a-distance in a big way in one initial instant accepted by most physicists, it hardly seems consistent to recoil at the thought of action at a distance.

After all, why should spatial separateness be elevated to the status of an absolute? After all, the universe is assumed by both physicists and philosophers to be one coherent whole where physical principles can be globally applied. Therefore already do physicists assume that the universe has global connections, albeit they are intellectual principles rather than material connections. But what justification do physicists have for their belief that physical principles are universal? And if there are such universal principles, do they not imply that the universe is an integrated whole before it is an aggregation of separated parts?

The doctrine, for that is what it is, that things spatially separated can have no influence on one another except by mechanistic touch, appears to need revision if the writings of Bohm [Boh93], Bell [Bel87] and Goldstein [Gol96B] are to be taken seriously. Interpreting Bell's work, Goldstein's argument [Gol96B] that non-locality is a necessary part of any quantum mechanical model is particularly strong. He points out that Bohm's hidden variable formulation is not an attempt to avoid non-locality; rather, it explicitly embraces non-locality, as much as any theory. In [Gol96A, Gol01] he successfully takes on the contentions of von Neumann [Neu32], Kochen-Specker [Koc67], and Wigner [Wig76] that no hidden variable theory of quantum mechanics is possible.

The universe might be taken as an initially undivided whole W that is thereafter subdivided by a spatial equivalence relation R into a set of equivalence classes {W/R}. These individuals (equivalence classes) can subsequently be aggregated into the spaces, atoms, molecules, planets and stars of our experience. Then, beside the relationships between nearby objects there could also be connections between non-local objects through their relationship to the whole, W. Admittedly this approach to a theory of everything is yet undeveloped, but it provides a philosophical basis for considering instantaneous action at a distance, not to mention other philosophical and physical motivations for such a theory.

As J. Raptis & R.R. Zapatrin [Rap02, p2] put it: "First comes the quantum, then space; not the other way around."

**1.13 Quantum Formalism**. In [Bal85] L.E. Ballentine provided a very clear account of conditional probability as it relates to the quantum mechanics formalism including a proof that the Kolmogorov axioms of probability are satisfied by this formalism.

"According to the standard postulate of quantum mechanics, the probability of obtaining the particular value $R = r_n$ is given by $P(R = r_n | \psi) = |\langle r_n | \psi \rangle|^2$, in the simplest case of a discrete nondegenerate eigenvalue spectrum and a pure state represented by the vector $\psi$." This is the "probability that the dynamical variable R has the value $r_n$ conditional on $\psi$." [Bal85, p885] He pointed out that the state $\Psi$ contains all the relevant information concerning the state preparation procedure.

Hardegree [Har76, pp56-58] provided a very clear exposition of the elements of quantum logic including an account of the "satisfaction function" h. This function assigns to each statement A (that some variable V has a value in the Borel set B) the set of all states (worlds) that verify A. This is the same "extension" function that appeared in [Cal87] and which formed the basis for assigning probabilities to propositions and conditional propositions in a way consistent with deductive logic.

Wilce [Wil02, pp18-19] has provided a good overview of quantum logic and probability including, of course, the orthodox quantum formalism: "Let H denote a complex Hilbert space and let $\mathcal{A}$ denote the collection of (unordered) orthonormal bases of H. Thus the outcome-space $X$ of $\cup \mathcal{A}$ will be the unit sphere of H. Note that if $u$ is any unit vector of H and $E \in \mathcal{A}$ is any orthonormal basis, we have

$$\sum_{x \in E} |\langle u, x \rangle|^2 = \| u \|^2 = 1.$$

Thus each unit vector of H determines a probability weight on $\mathcal{A}$. Quantum mechanics asks us to take this literally: any 'maximal' discrete quantum-mechanical observable is modeled by an orthonormal basis, and any pure quantum mechanical state, by a unit vector in exactly this way."

Concerning the Bohmian formulation of quantum mechanics, K. Berndl et al [Ber95] provided a technical but readable account including the use of a *conditional wave function* to deal with quantum subsystems from the hidden variable approach. This followed closely the account by D. Dürr et al [Dur95] going over similar territory with a somewhat less technical density.

**1.14 Interference**. As mentioned earlier, Ballentine [Bal86, p887] addressed the probability formulation of the double slit experiment and convincingly restated a serious objection raised by Koopman [Koo55] to Feynman's treatment of the experiment [Fey65, pp1-5]. Feynman claimed that since going through slit 1 and going through slit 2 are mutually exclusive events, the probabilities for the particle arriving at a position X when both slits are open should equal the sum of the probabilities of arriving at position X by going through the individual slits when one slit is blocked. Of course with both slits open some kind of interference occurs and so Feynman concluded that the probability law for the sum of two disjoint events is violated.

However, as Ballentine put it: "The probability of detection at X in the first case (only slit no. 1 open) should be written $P(X|C_1)$, where the conditional information $C_1$ includes (at least) the state function $\psi_1$ for the particle beam and the state $S_1$ (only slit no. 1 open). In the second case (only slit no. 2 open) the probability should be written as $P(X|C_2)$, where $C_2$ includes the state function $\psi_2$ and the screen state $S_2$ (only slit no. 2 open). In the third case (both slits open) the probability is of the form $P(X|C_3)$, where $C_3$ includes the state function $\psi_{12}$ (approximately equal to $\psi_1 + \psi_2$ but this fact plays no role in our argument) and the screen state $S_3$ (both slits open). We observe from experiment that $P(X|C_3) \neq P(X|C_1) + P(X|C_2)$." Ballentine went on to say that this fact, however, has no bearing on the validity of the sum rule for disjoint events of probability theory. As he said earlier, "…beware of probability statements expressed as $P(X)$ instead of $P(X|C)$. The second argument may be safely omitted only if the conditional event or information is clear from the context and *constant* throughout the problem. This is not the case in the double slit example."

On the other hand, accepting the Copenhagen interpretation of the two-slit experiment, S. Youssef [You91], [You94], You96], and [You01], showed that complex-valued probabilities can produce the interference results associated with that experiment. However, one must ask whether the probability function should be changed so much that we can no longer interpret it as a ratio? Should not a less radical extension be tried first? Shouldn't an extension of propositional logic to a non-distributive system of propositional fractions with associated standard conditional probabilities be tried before the values of the probability function become complex numbers?

**1.15 Superposition.**. Concerning superposition Ballentine [Bal85, p887] pointed out that the standard quantum mechanical formula $\langle B|A \rangle = \Sigma_C \langle B|C \rangle \langle C|A \rangle$ for amplitudes, where C is a set of non-overlapping sets connecting A and B, with the associated probabilities $P(B|A) = |\langle B|A \rangle|^2$, $P(B|C) = |\langle B|C \rangle|^2$ and $P(C|A) = |\langle C|A \rangle|^2$ is inconsistent with the probability formula $P(B|A) = \Sigma_C P(B|C)P(C|A)$ when the interference is not negligible. However that does not imply the superposition principle for amplitudes is inconsistent with classical probability theory because the most general probability formula for this situation is $P(B|A) = \Sigma_C P(B|C \wedge A)P(C|A)$, and the latter formula is not inconsistent with the amplitude formula. Ballentine concluded that the only remaining anomaly is that "joint probability distributions for two or more dynamical variables are not conventionally defined unless the corresponding operators are commutative."

**1.16 Entanglement and Quantum Computing.** While controversy still rages in the quantum research community about the truth of quantum non-locality and the reality of "quantum entanglement", the Defense Advanced Research Projects Agency (DARPA) of the U.S. Department of Defense is taking no chances about staying ahead of the rest of the world in this area of research. Soon after the appearance in 1994 of Shor's polynomial-time quantum computer algorithm for prime factorization [Sho99] DARPA initiated programs for research & development of quantum computers and associated algorithms. A. Steane [Ste97] gave a complete account of the field including classical information theory, experimental quantum information processors, and quantum error correction. A recent book by A. Pittenger [Pitt01] also nicely laid out the basics of quantum computing and error correcting. If quantum mechanics can be put into a simpler, near-Boolean framework, then these efforts to build quantum computers and design error correcting codes will be made easier, more intuitive and so save money.

**1.17 Summary.** In order to facilitate the use of Boolean fractions (conditional events) in expressing quantum relationships, Section 2 will lay down some stepping-stones for those who may be inclined to further investigate quantum mechanics in this framework. Hopefully these early results will be helpful to those who are in position to discover and express deeper quantum theorems with these methods. Therefore, several different approaches and concepts have been recast in terms of conditionals rather than an extensive investigation of any one of them. So this paper includes a survey of the various algebraic elements for quantum logic and how they can be translated into the conditional framework.

Section 2 starts with a careful review of the significant work of G. Schay on conditional event algebras including his choices of operations. Section 2.2 gives an account of the author's formulation of conditional event algebra (***B|B***) and the choice of operations. Section 2.3 recounts some relevant algebraic properties of (***B|B***). Theorem 2.4 and Corollary 2.5 are new, giving necessary and sufficient conditions for the two distributive laws for conditional events. Theorems 2.6 and Corollary 2.7 are new, characterizing the conditions for the two modular laws to hold for conditional events. Theorem 2.8 and Corollary 2.9 give new characterizations of the weak modularity laws for conditionals. Section 2.10 recounts the connection between three-valued logic and conditionals including the representation of conditionals as measurable indicator functions and the four 3-valued truth tables corresponding to the four operations on conditionals. Section 2.11 is a digression to motivate the need for the 3-valued, conditional event algebra even in ordinary chance situations. This section also serves to illustrate the non-monotonic aspects of conditionals in ordinary chance situations. Section 2.12 gives the superposition formulas for conditionals and suggests a correspondence with the linear forms of quantum logic. Section 2.13 is new and gives necessary and sufficient conditions for an extended additive law for conditional probability. Section 2.14 gives a preliminary analysis of simultaneous physical measurements in the context of conditionals. In sections 2.15-16 the quantum algebraic formulation of Varadarajan is introduced including orthogonality. A new definition of orthogonality for conditionals is given in Section 2.17 and the set of conditionals $(a|b)^\perp$ orthogonal to a given conditional $(a|b)$ is determined in Theorem 2.18. Theorem 2.19 shows that $(a|b)^\perp$ is closed under the conjunction and disjunction operations. In section 2.20 the conditional event counterparts to Varadarajan's orthocomplementation properties are listed. Motivated by Varadarajan's theorems, Section 3 on Quantum Logic and Conditional Events is totally new. Section 3.1 gives Varadarajan's convenient algebraic definition of simultaneous verifiability, and Theorem 3.2 gives new, intuitively clear necessary and sufficient conditions for simultaneous verifiability in the conditional event algebra. Corollary 3.3 gives another characterization of simultaneous verifiability for conditionals. In Definition 3.4 and Corollary 3.5 the notion of simultaneous falsifiability for conditionals is introduced and characterized. Corollary 3.6 proves that two conditionals are in a common Boolean sub-algebra just in case they are both simultaneous verifiable and falsifiable. Theorem 3.7 establishes that two conditionals (a|b) and (c|d) are in a common Boolean sub-algebra just in case b=d. Corollary 3.8 shows that two conditionals are simultaneous verifiable and falsifiable just in case they are in a common Boolean sub-algebra of (***B|B***). Theorem 3.9 establishes the uniqueness of the relative negation in (***B|B***). Section 3.10 gives the definition of an orthoalgebra as defined by Coecke et al, and Theorem 3.11 gives a counterpart for conditional events. Section 3.12 gives Piron's definition for an orthocomplementation on a lattice, and also a relative orthocomplementation. Next the Sasaki projection mappings are introduced in Section 3.13 according to equivalent accounts by Coecke, Delmas-Rigoutsos and Piron. These definitions are interpreted for conditionals in Definition 3.14, and then Theorem 3.15 and Corollary 3.16 give partial counterpart results for conditionals. It is made clear that conditionals events can distinguish relationships that are indistinguishable in the standard quantum formulation. Theorem 3.17 gives conditional counterparts to five other facts shown by Delmas-Rigoutsos for the Sasaki projection. Section 3.18 catalogues five different kinds of simultaneous verifiability distinguishable in the conditional event algebra. Section 3.19 discusses at least three different kinds of truth expressible with conditionals. Finally, suggesting future work, Section 3.20 relates quantum truth and quantum operations to some corresponding concepts in the conditional event realm including deductively closed sets of uncertain conditionals with respect to various natural deductive relations on conditionals.

**2. Boolean Fractions and Quantum Mechanics.** Concerning the philosophical basis for insisting on ordered pairs of propositions for representing information, I wish to state categorically that all information has explicit or implicit assumptions that must be represented in any adequate logic of information. Any so-called material fact is based upon the assumption of the validity of the human sensory system and related systems. This is assumed context, conditions for any and every material fact. Likewise every supposedly clear idea is founded upon assumptions of the validity of the human intellectual apparatus and conceptual framework. "I think" assumes that thought is real based upon nothing but personal human experience of what we all recognize as "thought". We assume that humans have common physical and mental experiences that can be mutually identified without further justification. But these constitute assumed conditions. And depending on the specific facts or ideas these conditions can easily vary.

So the overall premise of this paper is that the basic unit of information is a pair, not a single event or proposition. Every proposition or event has implicit assumptions, context. The basic unit of information must include a supposition (or condition) in the context of which another proposition or event is being identified.

Just as integer fractions dramatically improve the ability of integers to represent and operate on continuous distances in space, so too do propositional fractions dramatically improve the ability of propositions to represent and operate on uncertain conditional information in a situation space of variables taking on various possible values.

**2.1 The Conditional Event Algebras of Schay$^\Diamond$.** By 1968, G. Schay [Sha68] clearly had been developing his "algebra of conditional events" for quite some time before he published his one 10-page article. In it, he quickly defined a conditional event as a 3-valued indicator function as done here but without insisting that it be measurable. However his notation makes it clear that he was thinking of these indicator functions as defined on some kind of probability sample space $\Omega$. He defined two disjunction operations and two conjunction operations on pairs of these indicator functions as follows:

$$((A|B) \cap_s (C|D))(\omega) = \min\{(A|B)(\omega), (C|D)(\omega)\}, \text{ with domain } B \cap D, \quad (2.1)$$
$$((A|B) \cup_s (C|D))(\omega) = \max\{(A|B)(\omega), (C|D)(\omega)\}, \text{ with domain } B \cup D, \quad (2.2)$$
$$((A|B) \wedge_s (C|D))(\omega) = \min\{(A|B)(\omega), (C|D)(\omega)\}, \text{ with domain } B \cup D, \quad (2.3)$$
$$((A|B) \vee_s (C|D))(\omega) = \max\{(A|B)(\omega), (C|D)(\omega)\}, \text{ with domain } B \cap D, \quad (2.4)$$

and negation (~) on these indicator functions as

$$(\sim(A|B))(\omega) = (A'|B)(\omega), \quad (2.5)$$

where $A'$ is the negation or complement of A. (Schay used the over bar notation, $\bar{A}$, for the complement of A.) An "s" subscript is attached to Schay's operations to distinguish them from the operations to be recommended here.

---

$^\Diamond$ I wish to thank I.R. Goodman for discovering Schay's 1968 paper and bringing it to my attention.

Almost as an after thought, and without stating it is as a definition, Schay inserted in the first sentence of his Theorem 1, the equivalence relation for pairs of propositions:

$$(A|B) = (C|D) \text{ if and only if } B=D \text{ and } A \cap B = C \cap D. \quad (2.6)$$

He then pointed out (Theorem 1) that these operations on indicator functions can be expressed as operations on ordered pairs of propositions (conditional events) as follows:

$$(A|B) \cap_s (C|D) = (A \cap C \mid B \cap D), \quad (2.6)$$
$$(A|B) \cup_s (C|D) = ([(A \cup C) \cap B \cap D] \cup (A \cap B \cap D') \cup (B' \cap C \cap D) \mid B \cup D), \quad (2.7)$$
$$(A|B) \wedge_s (C|D) = ((A \cap B \cap C \cap D) \cup (A \cap B \cap D') \cup (B' \cap C \cap D) \mid B \cup D), \quad (2.8)$$
$$(A|B) \vee_s (C|D) = (A \cup C \mid B \cap D), \quad (2.9)$$
$$\sim(A|B) = (A'|B). \quad (2.10)$$

A typographic error in Schay's negation equation in Theorem 1 has mistakenly rendered it as $\sim(A|D) = (A'|B)$. In addition, the operation $\cup_s$ can obviously be simplified to

$$(A|B) \cup_s (C|D) = ((A \cap B) \cup (C \cap D) \mid B \cup D). \quad (2.11)$$

Concerning the second and third operations listed together with negation, Schay stated, 'As for the intuitive meaning of the operations, it seems that $\cup_s$, $\wedge_s$ and ~ correspond to the usual meaning of the words "or", "and" and "not".'

In fact, these are the very same, independently discovered operations on conditionals called "quasi" by Adams [Ada66], [Ada98, p164]. They were again independently discovered by Calabrese [Cal87] as part of a richer system with a fourth, iterated conditioning operation.

Later in his article (p343), Schay noted that these operations give meaning to expressions like $P((A|B) \cup_s (C|D))$, but that "such probabilities have strange properties". "It is simple", he said, "to construct examples such that $P(x \cup_s y) < P(x)$, or $P(x \wedge_s y) > 0$ although $P(y) = 0$, etc." These are the same non-monotonicities that prompted Adams to call these operations "quasi".

Schay also said that the usual rule $P(A|B)P(B) = P(A \cap B)$ can be dropped to handle special situations in which a change in condition "changes the results in an unusual way." His example was a 3-way election in which the probability of each candidate winning is 1/3, but if one candidate drops out, then one of the other two candidates gets all of his votes. Schay then said, "It may along these lines be possible to incorporate the probabilities of quantum mechanics in our theory". However, such probability examples as the vote problem above can be better handled by ordinary probability theory by defining more events to represent the situation. For instance, by defining the events "a", "b" and "c" of having each candidates in the race, the fact that A inherits all of C's votes can be expressed as $P(A|abc) = 1/3$ and $P(A|abc') = 2/3$, without violating standard conditional probability theory. So $P(A) = (1/3)P(abc) + (2/3)P(abc')$. Schay ended his paper suggesting connections be explored between his theory and quantum logic and quantum probability.

While ignoring the system with operations $\{\cup_s, \wedge_s, \sim\}$, which he said seems to correspond to the usual meanings of the words "or", "and" and "not", Schay focuses instead on two other combinations, one with operations $\cap_s$, $\cup_s$ and $\sim$, and the other with operations $\wedge_s$, $\vee_s$, and $\sim$. His Theorem 2 (p336) states that both of these systems of conditional events form distributive lattices. Schay probably had quantum mechanics in mind when he defined his two systems since the Hilbert space formulation of quantum mechanics starts with a (non-distributive) lattice. If so, the desire to form a lattice may have been misguided because they turned out to be distributive and so not applicable to quantum mechanics.

In an effort to extend the range of his operations to mappings such as $(B | A \cup B) \to (0|A)$ and $(B | A \cup B) \to (B|B)$, Schay defined four more operations. However, he did not define an iterated conditioning operation, which could have accomplished something like this in a less arbitrary way. For instance, using the iterated operation $((A|B) | C) = (A | B \cap C)$, which works for conditionals as well as simple Boolean propositions or events, and assuming A and B are disjoint (orthogonal) this mapping can be obtained by conditioning $(B | A \cup B)$ by $B'$, the negation of B:

$$((B | A \cup B) | B') = (B | (A \cup B) \cap B') = (B | A \cap B') = (0|A). \qquad (2.12)$$

Schay's main result (Theorem 5) is a generalization of Stone's well-known theorem [Sto36] characterizing any Boolean algebra as an algebra of subsets under union, intersection and complement.

**2.2 Conditional Event Algebra.** The conditional event algebra introduced by Calabrese in 1987 has been thoroughly developed ([Cal87], [Cal91], [Cal94]) including deduction of uncertain conditionals [Cal02]. It is completely consistent with standard conditional probability theory but extends it to 3-valued Boolean fractions (conditional events), which are "undefined" in case their condition is false.

Let $(\mathcal{B}|\mathcal{B})$ denote the set of ordered pairs, $\{(a|b): a, b \text{ in } \mathcal{B}\}$, called the set of *conditionals*, "a given b", of $\mathcal{B}$. The proposition or event "b" is called the *condition, premise* or *antecedent* and the proposition or event "a" is called the *consequent* or *conclusion*. Define two conditionals to be equivalent (=) as in equation (2.6): (a|b) = (c|d) if and only if b=d and ab = cd.

The algebra of Boolean fractions provides four operations for "or", "and", "not" and also "given" (conditioning) to handle compound conditional forms such as "((a|b) | (c|d))".

**2.2.1 Relative Negation**. The relative negation of "a given b" is the "negation of a, given b". That is,

$$(a|b)' = (a' | b), \qquad (2.13)$$

and the latter has probability 1 - P(a|b).

**2.2.2 Disjunction.** Concerning disjunction, "if b then a, or if d then c" means "if either conditional is applicable then at least one is true". That is,

$$(a|b) \vee (c|d) = (ab \vee cd) | (b \vee d) \qquad (2.14)$$

**2.2.3 Conjunction.** Concerning conjunction, "if b then a and if d then c" means "if either conditional is applicable then one is true while the other is not false". That is,

$$(a|b) \wedge (c|d) = [ab(c \vee d') \vee (a \vee b')cd] | (b \vee d) \qquad (2.15)$$
$$= (abd' \vee abcd \vee b'cd | b \vee d), \qquad (2.16)$$

which also means "if either conditional is applicable then either they are both true or else one is true while the other is inapplicable."

**2.2.4 Iterated Conditioning.** A conditional (c|d) may itself be a condition for another proposition or conditional proposition; a fraction can be in the denominator.

$$((a|b) | (c|d)) = (a | b \wedge (c|d)) \qquad (2.17)$$
$$= (a | b (c \vee d')). \qquad (2.18)$$

The order of preference of the operations is negation ( ' ) before conjunction, before disjunction, before conditioning ( | ).

The algebra (𝓑 | 𝓑) of conditionals includes the original Boolean algebra 𝓑 as those conditionals (a|Ω), where Ω is the universal event, and "a" is any member of 𝓑. In logical notation these are the conditionals (a|1) whose condition is certain. Analogously, these are like the integer fractions whose denominators are 1. Fixing the condition b also yields a Boolean algebra (𝓑|b) = {(a|b): all a ∈ 𝓑 }.

Among the properties retained by the algebra of conditionals are the two familiar de Morgan formulas by which conjunction can be expressed in terms of disjunction and negation, or disjunction can be expressed in terms of negation and conjunction.

**2.3 Properties of Conditional Event Algebra.** The properties of this conditional event algebra have been listed in [Cal87] and completely characterized in [Cal02]. The disjunction and conjunction operations are associative, commutative and idempotent. The zero conditional (0|1) is an absolute zero with respect to conjunction in the sense that (a|b) ∧ (0|1) = (0|1), but (a|b) ∨ (0|1) = ab rather than (a|b). Similarly, (1|1) is an absolute unity element with respect to disjunction in the sense that (a|b) ∨ (1|1) = (1|1), but (a|b) ∧ (1|1) = a ∨ b' rather than (a|b). However, any conditional (a|b) has a relative complement (a'|b) such that (a|b) ∨ (a'|b) = (1|b) and (a|b) ∧ (a'|b) = (0|b). Negation also satisfies the law of double negation: ((a|b)')' = (a|b).

In Boolean algebra the equation B ∧ A = B ∧ (A|B) always holds. This also holds with conditionals: (a|b) ∧ (c|d) = (c|d) ∧ [(a|b) | (c|d)] by expanding the right hand side using the operations on conditionals. Concerning distributivity there is the following result.

**2.4 Theorem on Distributivity.** (a|b) ∧ [(c|d) ∨ (e|f)] = (a|b)(c|d) ∨ (a|b)(e|f) if and only if (ab)(e'f) ≤ d and (ab)(c'd) ≤ f. That is, conjunction distributes over disjunction just in case the truth of the outside conditional and the falsity of one of the inside conditionals implies the other inside conditional is applicable.

**Proof of Theorem 2.4.** [(c|d) ∨ (e|f)] = (cd ∨ ef | d ∨ f). Conjoining this with (a|b) yields

$$(abd'f' \vee (b' \vee ab)(cd \vee ef) \mid b \vee d \vee f).$$

On the other hand by expanding and collecting terms it follows that (a|b)(c|d) ∨ (a|b)(e|f) =

$$(abd'f' \vee abd'f \vee abdf' \vee (b' \vee ab)(cd \vee ef) \mid b \vee d \vee f).$$

These two conditionals will be equal if and only if abd'f ∨ abdf' ≤ abd'f' ∨ (b' ∨ ab)(cd ∨ ef). Since (abd'f') is disjoint from (abd'f ∨ abdf'), this is equivalent to

$$abd'f \vee abdf' \leq (b' \vee ab)(cd \vee ef),$$

and since = b'(cd ∨ ef) is disjoint from (abd'f ∨ abdf'), the inequality is equivalent to

$$abd'f \vee abdf' \leq abcd \vee abef.$$

Since abd'f is disjoint from abcd, therefore abd'f ≤ abef. Similarly, abdf' ≤ abcd. On the other hand, if these latter two inequalities hold then the one above obviously holds. Finally, abd'f ≤ abef is equivalent to abd'fe' = 0, which becomes (ab)(e'f) ≤ d, one of the two inequalities to be proved. Similarly abdf' ≤ abcd is equivalent to (ab)(c'd)f' = 0, which becomes (ab)(c'd) ≤ f, the other inequality to be proved. That completes the proof of the theorem.

**2.5 Corollary on Distributivity.** Distributivity of disjunction over conjunction. (a|b) ∨ [(c|d) ∧ (e|f)] = [(a|b) ∨ (c|d)] ∧ [(a|b) ∨ (e|f)] if and only if (a'b)(ef) ≤ d and (a'b)(cd) ≤ f. That is, disjunction distributes over conjunction if and only if whenever the outside conditional is false and one inside conditional is true then the other inside conditional is applicable.

**Proof of Corollary 2.5.** Using the de Morgan formula,

$$(a|b) \vee [(c|d) \wedge (e|f)] = \{(a|b)' \wedge [(c|d) \wedge (e|f)]'\}'$$
$$= \{(a'|b) \wedge [(c|d)' \vee (e|f)']\}'$$
$$= \{(a'|b) \wedge [(c'|d) \vee (e'|f)]\}'$$
$$= \{[(a'|b) \wedge (c'|d)] \vee [(a'|b) \wedge (e'|f)]\}',$$

where by the theorem the last equality is true if and only if $(a'b)(ef) \leq d$ and $(a'b)(cd) \leq f$. But

$$\{[(a'|b) \wedge (c'|d)] \vee [(a'|b) \wedge (e'|f)]\}' = [(a'|b) \wedge (c'|d)]' \wedge [(a'|b) \wedge (e'|f)]'$$
$$= [(a'|b)' \vee (c'|d)'] \wedge [(a'|b)' \vee (e'|f)']$$
$$= [(a|b) \vee (c|d)] \wedge [(a|b) \vee (e|f)],$$

and that completes the proof of the corollary.

As part of the algebraic formulation of quantum logic Birkhoff and von Neumann defined a *modular* lattice [Bir36, p833] as one for which $A \cup [B \cap C] = [A \cup B] \cap C$ whenever $A \leq C$. For conditionals, there is the following counterpart:

**2.6 Theorem on Modular Law.** Three conditionals $(a|b)$, $(c|d)$ and $(e|f)$ satisfy the disjunction-over-conjunction modular law,

$$(a|b) \vee [(c|d) \wedge (e|f)] = [(a|b) \vee (c|d)] \wedge (e|f)$$

if and only if $(ab)(e'f) = 0$ and $(a'b)(ef) \leq d$. That is, the modular law holds if and only if

1) the truth of $(a|b)$ is inconsistent with the falsity of $(e|f)$, and
2) if $(a|b)$ is false and $(e|f)$ is true then $(c|d)$ is applicable.

**Proof of Theorem 2.6.** By expanding both sides of the modular law and equating conditionals it follows that

$$ab \vee d'ef \vee cd(f' \vee ef) = ab(f' \vee ef) \vee cd(f' \vee ef) \vee b'd'ef.$$

Conjunction on both sides by $e'f$, the complement of $(f' \vee ef)$, yields $(ab)(e'f) = 0$. That is, $ab \leq (f' \vee ef)$. Conjunction on both sides by $bd'$ yields $abd' \vee bd'ef = ab(f' \vee ef)d' = abd'$. So $bd'ef \leq abd'$. That is, $bd'ef(abd')' = 0$. That is, $a'bd'ef = 0$. That is, $(a'b)(ef) \leq d$. That establishes the necessity of 1) and 2) of the theorem. For sufficiency, note that if both conditions 1) and 2) hold, then $ab(f' \vee ef) = ab$ and $bd'ef \leq ab$. So the right hand side of the equation reduces to the left hand side. That completes the proof of the theorem.

**2.7 Corollary on Modular Law.** Three conditionals $(a|b)$, $(c|d)$ and $(e|f)$ satisfy the conjunction-over-disjunction modular law,

$$(a|b) \wedge [(c|d) \vee (e|f)] \ = \ [(a|b) \wedge (c|d)] \vee (e|f)$$

if and only if $(a'b)(ef) = 0$ and $(ab)(e'f) \leq d$. The first of these is equivalent to $ef \leq a \vee b'$, that the truth of $(e|f)$ implies the non-falsity of $(a|b)$. The second inequality says the truth of $(a|b)$ and the falsity of $(e|f)$ implies the applicability of $(c|d)$.

**Proof of Corollary 2.7.** By the theorem applied to the negations of the three conditionals,

$$(a'|b) \vee [(c'|d) \wedge (e'|f)] \ = \ [(a'|b) \vee (c'|d)] \wedge (e'|f)$$

if and only if $(a'b)(ef) = 0$ and $(ab)(e'f) \leq d$. Using both de Morgan formulas on both sides yields the required conjunction-over-disjunction modular law.

In the theorem or corollary above, if $(c|d) = (a|b)'$ then $b=d$, and so $(ab)(e'f) \leq d$ and $(a'b)(ef) \leq d$ are always satisfied leaving just the condition $(ab)(e'f) = 0$ for the theorem and $(a'b)(ef) = 0$ for the corollary.

**2.8 Corollary on Weak Modularity.** If $(a|b)$ and $(e|f)$ are two conditionals then $(a|b) \wedge [(a|b)' \vee (e|f)] = (e|f)$ if and only if $b \leq f$ and $a'b \leq e'f$. These conditions mean that the applicability of $(a|b)$ implies the applicability of $(e|f)$ and the falsity of $(a|b)$ implies the falsity of $(e|f)$.

**Proof of Corollary 2.8.** By the remarks above concerning $(c|d) = (a|b)'$, the first corollary of the theorem yields that $(a|b) \wedge [(a|b)' \vee (e|f)] = [(a|b) \wedge (a|b)'] \vee (e|f)$ if and only if $(a'b)(ef) = 0$. This is equivalent to $a'b \leq e'f \vee f'$. But $[(a|b) \wedge (a|b)'] \vee (e|f) = (0|b) \vee (e|f) = (ef \mid b \vee f) = (e|f)$ if and only if $(b \vee f = f)$. This is equivalent to $b \leq f$ and so to $f' \leq b'$. Therefore $a'b \leq e'f \vee f' \leq e'f \vee b'$, and so $(a'b \leq e'f \vee b')$ yields $(a'b \leq e'f \ b \leq e'f)$ by conjoining both sides by b. That completes the proof of the corollary.

**2.9 Corollary on Weak Dual-Modularity.** If $(a|b)$ and $(e|f)$ are two conditionals then $(a|b) \vee [(a|b)' \wedge (e|f)] = (e|f)$ if and only if $b \leq f$ and $ab \leq ef$. These conditions mean that the applicability of $(a|b)$ implies the applicability of $(e|f)$ and the truth of $(a|b)$ implies the truth of $(e|f)$.

**Proof of Corollary 2.9.** Expansion of the equation yields that $(ab \vee a'bf' \vee b'ef \vee a'bef \mid b \vee f) = (e|f)$. So $ab \vee a'bf' \vee b'ef \vee a'bef = ef$ and $b \vee f = f$. The latter equation is equivalent to $(b \leq f)$ and to $bf' = 0$. So the left side of the first equation reduces to $ab \vee b'ef \vee a'bef = ab \vee (b' \vee a'b)ef = ab \vee (ab)'ef = ab \vee ef$. So the first equation is equivalent to $ab \vee ef = ef$. That is, $ab \leq ef$. That completes the proof.

For two conditionals $(a|b)$ and $(e|f)$ the relations $b \leq f$ and $ab \leq ef$ were defined [Cal02] as $(a|b) \leq_{ap} (e|f)$ and $(a|b) \leq_{tr} (e|f)$ respectively, and their combination defined as the deductive relation $(a|b) \leq_\vee (e|f)$.

**2.10 Three-Valued Logic**. Conditionals (Boolean fractions) can be represented by 3-valued measurable indicator functions [Cal91, p76]: Suppose that two measurable Boolean propositions "a" and "b" are represented by their associated measurable indicator functions:

$$a(\omega) = \begin{cases} 1, & \text{if a is true in state } \omega, \\ 0, & \text{if a is false in state } \omega, \end{cases} \quad \text{and} \quad b(\omega) = \begin{cases} 1, & \text{if b is true in state } \omega, \\ 0, & \text{if b is false in state } \omega. \end{cases}$$

In this formulation a Boolean fraction or conditional event is an indicator function (a|b) given by:

$$(a|b)(\omega) = \begin{cases} 1, & \text{if a and b are true in } \omega, \\ 0, & \text{if a is false and b is true in } \omega, \\ \text{Undefined}, & \text{if b is false in } \omega. \end{cases}$$

Each operation on conditionals easily generates a 3-valued truth table by setting a, b, c, and d equal to 1 or 0 in equations (2.13), (2.14), (2.16) and (2.18). The converse is also true. See [Cal93, p7]. The four 3-valued truth tables for conditionals are listed in the following table:

|   | AND |   |   | OR |   |   | GIVEN |   |   | NOT |
|---|---|---|---|---|---|---|---|---|---|---|
|   | T | F | U | T | F | U | T | F | U |   |
| T | T | F | T | T | T | T | T | U | T | F |
| F | F | F | F | T | F | F | F | U | F | T |
| U | T | F | U | T | F | U | U | U | U | U |

**Table 1. Three-Valued Truth Tables for Conditional Propositions**

**2.11 Three-Valued (Conditional) Computations.** While providing a clear and elementary account of the mathematics of the quantum formalism A. Landé [Lan76, p436] said that "an artificial difference between the classical and quantum realms has been established by those who think that macroscopic physics can be understood in terms of ordinary logic, but that atomic physics requires the introduction of a *three-valued logic*, true, false and undetermined." "We do not deny", he says, "anyone the pleasure of this mental gymnastics. But why only in case of the quantum realm when this logic fits (or does not fit) ordinary chance situations as well!"

My response to this rhetorical challenge is to assert that mathematicians and physicists have in fact for a long time been ignoring the need for such a 3-valued logic in ordinary chance situations!

To demonstrate this little recognized fact, consider the following game of chance. An ordinary six-sided die with faces numbered from 1 through 6 is rolled once. Someone makes the following statement:

> If the roll is an even number then it will be a "2" or if the roll is "less than 5" then it will be "less than 4".

What is the probability of winning this bet? More importantly why is there no standard way of calculating this probability in terms of the component probabilities and conditional probabilities? Why is there no standard way of operating with the conditional events? Some people even claim erroneously that such compound conditional statements have no meaning, but there are all sorts of examples of these kinds of statements in natural language. (One need only say something like "if the store has soda then buy me some soda, and if the store has beer then buy me some beer" to have an example of a compound conditional English sentence that makes perfect sense.) See for instance [Cal96, pp192-203] for several practical problems completely solved using this conditional event algebra.

The answer to this die problem above can be determined by brute force examination of each possible outcome from 1 to 6 to see if it first satisfies the condition and then whether it satisfies the consequent. In this way the outcomes 1, 2, 3, 4 and 6 are identified as satisfying one or both of the conditions of being "even" or "less than 5". Of these five outcomes, 4 or 6 will lose the bet while 1, 2, or 3 will win. So the probability of winning is 3/5. If outcome 5 turns up then the bet is neither won nor lost. This is the inapplicable case where the $3^{rd}$ truth-value comes in.

But this probability could also have been calculated using the 3-valued operations on conditionals in the system of Boolean fractions as follows:

$$(2 \mid even) \vee (< 4 \mid < 5)$$
$$= (\{(2) \cap even\} \cup \{(< 4) \cap (< 5)\}) \mid (even \cup (< 5))$$
$$= (\{2, 1, 2, 3\} \mid \{1, 2, 3, 4, 6\})$$
$$= (\{1, 2, 3\} \mid \{1, 2, 3, 4, 6\}),$$

which has conditional probability 3/5. In more complicated situations the answer can be calculated using the general formula [Cal91, p84]

$$P((a|b) \vee (c|d)) = P(a|b)P(b| b \vee d) + P(c|d)P(d| b \vee d) - P(abcd \mid bd)P(bd \mid b \vee d)$$
$$= (1/3)(3/5) + (3/4)(4/5) - (1/2)(2/5) = 3/5.$$

where a = (2), b = (even), c = (<4) and d = (<5), and juxtaposition has replaced conjunction $\wedge$.

Notice that this "superposition" (disjunction) of two conditionals (a|b) and (c|d) that individually have conditional probabilities 1/3 and 3/4 respectively, has a combined probability value of 3/5, which is *between* 1/3 and 3/4, not above them both as is always true for Boolean propositions.

Taken in the abstract this feature of the algebra of conditionals seems unintuitive and hard to swallow. But it clearly happens in a simple die throw when two propositions with different conditions are disjoined. The reason for the possibility of non-monotonicity when conditionals are combined is due to the possibilities of an expanded context (condition).

For instance, were someone to bet "if the die role is even then it will be a 2, 4, or 6", they would win with probability 1. But suppose the person said instead, "If the die role is even then it will be a 2, 4, or 6, or if the roll is odd then it will be a 5." Here, since the context (condition) has expended to all 6 outcomes, and because only 4 of them are winning outcomes, the probability of winning has gone down from 1 to 2/3 even though the person used "or" between the two clauses.

This is simply the non-monotonic way conditionals with different conditions operate. It is also easy to construct such simple examples in which $P((a|b) \wedge (c|d)) > P(a|b)$.

**2.12 Disjunction and Conjunction Superposition Formulas.** As stated in [Cal94, p1682], without further qualification, for any conditionals (a|b) & (c|d),

$$(a|b) \vee (c|d) = (a|b)(b| \, b \vee d) \vee (c|d)(d| \, b \vee d).$$

This can easily be verified by applying the operations on conditionals to see that the two sides are equal. It is also easy to verify that

$$(a|b) \vee (c|d) = (a|b)(bd'| \, b \vee d) \vee (c|d)(b'd| \, b \vee d) \vee ((a \vee c)bd \, | \, b \vee d),$$
$$P((a|b) \vee (c|d)) = P(a|b)P(bd'| \, b \vee d) + P(c|d)P(b'd| \, b \vee d) + P((a \vee c)bd \, | \, b \vee d).$$

Similarly,

$$(a|b) \wedge (c|d) = (a|b)(bd'| \, b \vee d) \vee (c|d)(b'd| \, b \vee d) \vee ((a \wedge c)bd \, | \, b \vee d),$$
$$P((a|b) \wedge (c|d)) = P(a|b)P(bd'| \, b \vee d) + P(c|d)P(b'd| \, b \vee d) + P((a \wedge c)bd \, | \, b \vee d).$$

Notice also that that if (a|b) & (c|d) are two conditionals for which the truth of one implies the non-falsity of the other, that is, if $(ab)(c'd) = 0 = (a'b)(cd)$, like when b and d are disjoint, then

$$(a|b) \vee (c|d) = (a|b) \wedge (c|d),$$

(and conversely) and so their probabilities are equal. In this case the last term in both of the above probability formulas becomes $P(abcd \, | \, b \vee d)$. If b and d are disjoint then the probability of both $(a|b) \vee (c|d)$ and $(a|b) \wedge (c|d)$ is $P(a|b)P(b \, | \, b \vee d) + P(c|d)P(d \, | \, b \vee d)$. This is a weighted average of $P(a|b)$ and $P(c|d)$ with weights $P(b \, | \, b \vee d)$ and $P(d \, | \, b \vee d)$ respectively expressing the relative probabilities of b and of d given that either is true. If also a=c, then

$$(a \, | \, b \vee d) = (a|b)(b \, | \, b \vee d) \vee (a|d)(d \, | \, b \vee d),$$
$$P(a \, | \, b \vee d) = P(a|b)P(b \, | \, b \vee d) + P(a|d)P(d \, | \, b \vee d).$$

Recalling the quantum formalism, note how $[(b| \, b \vee d), (d| \, b \vee d)]$ is a unit vector representing a "state", and $[(a|b), (a|d)]$ is some kind of projection (conditioning) of "a" onto b and onto d. In general this is just

$$P(a|u) = \Sigma_i \, P(a|u_i)P(u_i|u)$$

Concerning under what circumstances the conditional probability function is additive there is the following result:

**2.13 Theorem on the Additive Law.** $P((A|C_1) \vee (B|C_2)) = P(A|C_1) + P(B|C_2)$ if and only if one of the following is true:

1) $P(A|C_1) = 0 = P(B|C_2)$,
2) $P(A|C_1) = 0$ and $C_1 \leq C_2$ a.s.,
3) $P(A|C_2) = 0$ and $C_2 \leq C_1$ a.s.,
4) $C_1 = C_2$ a.s. and $P(A \wedge B \, | \, C_1) = 0$.

**Proof of Theorem 2.13**: In general

$$P((A|C_1) \vee (B|C_2)) = P(A|C_1)P(C_1| C_1 \vee C_2) + P(B|C_2)P(C_2| C_1 \vee C_2) - P(AC_1BC_2 | C_1 \vee C_2).$$

If this equals $P(A|C_1) + P(B|C_2)$, rearranging terms yields

$$P(A|C_1)[1 - P(C_1| C_1 \vee C_2)] + P(B|C_2)[1 - P(C_2| C_1 \vee C_2)] = -P(AC_1BC_2 | C_1 \vee C_2).$$

Since all terms on the left hand side are non-negative, but the right hand side is non-positive, all terms must be zero. So $P(A|C_1)[1 - P(C_1| C_1 \vee C_2)] = 0$, $P(B|C_2)[1 - P(C_2| C_1 \vee C_2)] = 0$, and $P(AC_1BC_2 | C_1 \vee C_2) = 0$. From these 3 equations the result easily follows by cases.

Thus only when the conditions $C_1$ and $C_2$ are equal almost surely is there a non-trivial realization of the sum formula for disjoint events A and B in the context of the common condition $C_1$.

P. Busch [Bus96, p2] noted that when $\psi$ is a proper superposition of eigenstates of a quantum observable B, then B is indeterminate in the state, $\psi$. That is, B does not have a definite value. This corresponds to the Hilbert space projection operator T for B satisfying neither $T\psi = \psi$ nor $T\psi = 0$. Unlike Boolean algebra, the system of Boolean fractions naturally includes conditional propositions B whose value $B(\omega)$ for a given state $\omega$, can be 1 or 0 or "undefined".

In a short paper P. Busch [Bus98, p1] observed, "it is possible to formulate conditions (such as restrictions of the set of states, or superselection rules) under which a quantum system will appear to display (approximately) classical behavior. But a theoretical explanation of why, and under what circumstances, such classical conditions come to be satisfied seems to be lacking." Perhaps Boolean fractions can illuminate this issue since for any fixed condition the system is Boolean, but fractions with different conditions always form a non-Boolean subsystem.

**2.14 Simultaneous Physical Measurements.** To see how conditionals might represent the subtleties of simultaneous measurements, suppose b is a proposition or event describing the experimental conditions and apparatus for measuring the position of a particle q, and let a be the proposition or event representing the act of measuring the position of q. So (a|b) is the conditional referring to the position measurement of q given the experimental conditions for measuring that position. Similarly, let (c|d) be the conditional representing the act c of measuring the velocity of q given the experimental conditions d for measuring the velocity of q. Therefore if we wish to accomplish both a measurement of position and of velocity under the appropriate conditions, we want $(a|b) \wedge (c|d)$. Now suppose there were a state or condition s such that $b \wedge s = d \wedge s \neq 0$. That is, suppose $(b|s) = (d|s) \neq (0|s)$. In other words, suppose that given state s, $b = d \neq 0$. Then conditioning $(a|b) \wedge (c|d)$ by s easily yields $((a|b) \wedge (c|d) | s) = ((a|b)|s) \wedge ((c|d)|s) = (a|bs) \wedge (c|ds) = (a|bs) \wedge (c|bs) = (a \wedge c | bs)$, indicating that a and c can be measured under the joint condition $b \wedge s$.

However if there is no such state s, that is, if $b \wedge d = 0$, then in any state s in which condition b is satisfied ($s \leq b$) condition d cannot be satisfied because $s \wedge d \leq b \wedge d = 0$. So measuring position yields $((a|b) \wedge (c|d) | b) = ((a|b)|b) \wedge ((c|d)| b) = (a|b) \wedge (c|db) = (a|b) \wedge U = (a|b)$. And measuring velocity similarly yields (c|d). But trying to measure both position and velocity by satisfying both measurement conditions yields ((a|b) ∧ (c|d) | bd) = (a|bd) ∧ (c|bd) = U ∧ U = U. One or the other condition, b or d, can first be imposed on (a|b) ∧ (c|d) to get a measurement, but as soon as one condition is satisfied the other can't be satisfied. This is the source of non-commutativity and non-distributivity of quantum measurements.

**2.15 Quantum Logic.** Varadarajan [Var68, p105] adopted the structure of an orthocomplemented lattice as the basic algebraic construct for quantum logic as follows:

**2.16 Definition of Orthocomplementation.** An orthocomplementation of a lattice $\mathcal{L}$ under partial order < is a mapping (⊥): a → $a^\perp$ of $\mathcal{L}$ into $\mathcal{L}$ such that

   i)    ⊥ is one-to-one and maps $\mathcal{L}$ onto itself,
   ii)   a < b implies $b^\perp$ < $a^\perp$,
   iii)  $a^{\perp\perp}$ = a for all a,
   iv)   a ∧ $a^\perp$ = 0  for all a,
   v)    a ∨ $a^\perp$ = 1  for all a.

Varadarajan noted that (ii)-(iv) imply (i) and (v).

**2.17 Definition of Orthogonality for Conditionals.** Two conditionals (a|b) and (c|d) are said to be *orthogonal* (disjoint) if (a|b) ∧ (c|d) = (0| b ∨ d). This will be expressed as (a|b) ⊥ (c|d). The set $(a|b)^\perp$ = {(c|d): (c|d) ⊥ (a|b)} is the set of all conditionals orthogonal to (a|b).

Expressed in terms of the components of (a|b) and (c|d) this is easily equivalent to having ab ≤ c′d and cd ≤ a′b. (0 = abd′ ∨ b′cd ∨ abcd = ab(d′ ∨ cd) ∨ (b′ ∨ ab)cd ⇔ ab(d′ ∨ cd) = 0 and ab(d′ ∨ cd) = 0. Therefore ab ≤ (d′ ∨ cd)′ = c′d, and similarly for cd ≤ a′b.)

In [Cal02] (and earlier too) the deductive relation $\leq_{pm}$ was defined, and it was shown that (a|b) $\leq_{pm}$ (c|d) just in case ab ≤ cd and c′d ≤ a′b. So (a|b) and (c|d) are orthogonal in case (a|b) $\leq_{pm}$ (c|d)′. ["pm" refers to "probabilistically monotonic".] The relation ab ≤ cd has been denoted (a|b) $\leq_{tr}$ (c|d), meaning the truth of (a|b) implies the truth of (c|d). The relation c′d ≤ a′b is equivalent to a ∨ b′ ≤ c ∨ d′, which has been referred to as the deductive relation (a|b) $\leq_{nf}$ (c|d), meaning the non-falsity of (a|b) implies the non-falsity of (c|d).

**2.18 Theorem on Orthogonality.** The set of conditionals orthogonal to (a|b) is $(a|b)^\perp$ = {(a′bx | ab ∨ y): x, y in $\mathcal{B}$}.

**Proof of Theorem 2.18**. $(a|b)^\perp$ = {(c|d): (c|d) ⊥ (a|b)} = {(c|d): (a|b) ∧ (c|d) = (0|b ∨ d)}. So (abd′ ∨ b′cd ∨ abcd = 0. Therefore abd′ = 0, b′cd = 0, and abcd = 0. So ab ≤ d, cd ≤ b, cd ≤ (ab)′. Combining the last two inequalities yields cd ≤ b(ab)′ = a′b, which implies both of them. So cd = a′bx, for some x in $\mathcal{B}$ and d = ab ∨ y, for some y in $\mathcal{B}$. So (c|d) = (a′bx | ab ∨ y) for some x, y in $\mathcal{B}$. That is, $(a|b)^\perp$ = {(a′bx | ab ∨ y): x, y in $\mathcal{B}$}. That completes the proof of the theorem.

**2.19 Theorem on Orthogonal Closure.** The set $(a|b)^\perp = \{(c|d): (c|d) \perp (a|b)\}$ of all conditionals orthogonal to $(a|b)$ is closed under disjunction and conjunction.

**Proof of Theorem 2.19.** If $(c|d) \in (a|b)^\perp$ and $(e|f) \in (a|b)^\perp$ then $(c|d) \vee (e|f) = (cd \vee ef \mid d \vee f) \in (a|b)^\perp$ just in case $cd \vee ef \leq a'b$ and $ab \leq (cd \vee ef)'(d \vee f)$. Since $(c|d) \in (a|b)^\perp$ therefore $cd \leq a'b$ and $ab \leq c'd$. Similarly $ef \leq a'b$ and $ab \leq e'f$. So $cd \vee ef \leq a'b$ and $ab \leq (c'd)(e'f) \leq (cd \vee ef)'(d \vee f)$. Therefore $[(c|d) \vee (e|f)] \perp (a|b)$. The conjunction $(c|d) \wedge (e|f) = (c \vee d')(e \vee f') \mid (d \vee f) = (cdf' \vee d'ef \vee cdef \mid d \vee f) \in (a|b)^\perp$ just in case $cdf' \vee d'ef \vee cdef \leq a'b$ and $ab \leq [(c \vee d')(e \vee f')]'(d \vee f)$. Since $cd \leq a'b$ and $ef \leq a'b$ therefore $cdf' \vee d'ef \vee cdef \leq a'b$. Now $[(c \vee d')(e \vee f')]'(d \vee f) = [(c \vee d')' \vee (e \vee f')'](d \vee f) = c'd \vee e'f$. Since $ab \leq c'd$, therefore $ab \leq (c'd) \vee (e'f)$. Therefore $[(c|d) \wedge (e|f)] \perp (a|b)$. That completes the proof.

If A and B are two Boolean events and $AB = 0 = A'B'$ then $B = A'$. Similarly, it is easy to show that if $(a|b) \wedge (c|d) = (0 \mid b \vee d) = (a|b)' \wedge (c|d)'$ then $(c|d) = (a|b)'$. Note that just as $A'B' = 0$ means $A \vee B = 1$, for conditionals, $(a|b)' \wedge (c|d)' = (0 \mid b \vee d)$ means $(a|b) \vee (c|d) = (1 \mid b \vee d)$.

Now, although the algebra of Boolean fractions under the disjunction, conjunction and negation operations of Section 2.2 does not form a lattice, it almost does, and having a full lattice does not seem to be necessary to capture the full quantum dynamics. Concerning the algebra $(\mathcal{B}|\mathcal{B})$ of Boolean fractions of a Boolean algebra $\mathcal{B}$ there is the following result that parallels Varadarajan:

**2.20 Complementation in $(\mathcal{B}|\mathcal{B})$.** A conditional event algebra $(\mathcal{B}|\mathcal{B})$ has a relative complement $(a|b)'$ for each member $(a|b)$ in $(\mathcal{B}|\mathcal{B})$ such that

1) $( ' )$ is one-to-one and onto $(\mathcal{B}|\mathcal{B})$,
2) $(a|b) \leq_{pm} (c|d) \Rightarrow (c|d)' \leq_{pm} (a|b)'$,
3) $((a|b)')' = (a|b)$,
4) $(a|b) \wedge (a|b)' = (0|b)$ for all $(a|b)$ in $(\mathcal{B}|\mathcal{B})$,
5) $(a|b) \vee (a|b)' = (1|b)$ for all $(a|b)$ in $(\mathcal{B}|\mathcal{B})$.

Using the operations on conditionals these are all very easy to show.

## 3. Quantum Logic & Conditional Events.
From a purely logical point of view, some pairs of quantum events or pairs of quantum propositions cannot be simultaneously verified. This notion of simultaneous verifiability has been conveniently expressed algebraically by V.S. Varadarajan [Var68, p118]:

**3.1 Definition of Simultaneous Verifiability.** Two members "a" and "c" of a logic $\mathcal{L}$ are *simultaneously verifiable*, $a \leftrightarrow c$, if there are members $a_1$, $c_1$, and $e$ of $\mathcal{L}$ such that $a_1 \wedge c_1 = 0$ and $a_1 \wedge e = 0$ and $e \wedge c_1 = 0$, and with $a = a_1 \vee e$ and $c = e \vee c_1$.

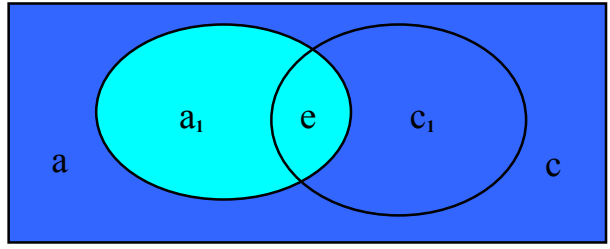

**Figure 1. Simultaneous Verifiability in an Orthocomplemented Lattice**

Obviously, in a Boolean algebra any two members "a" and "c" are always simultaneously verifiable. However in conditional event algebra only some conditionals are simultaneously verifiable:

**3.2 Theorem on Simultaneous Verifiability.** In the conditional event algebra two conditionals $(a|b)$ and $(c|d)$ are *simultaneously verifiable*, $(a|b) \leftrightarrow (c|d)$, if and only if $ab \leq d$ and $cd \leq b$.

That is, two conditionals are simultaneously verifiable if and only if the truth of one conditional implies the applicability of the other conditional. That is, two conditionals are simultaneously verifiable if verifying one implies that the other is applicable and so can be verified.

**Proof of Theorem 3.2.** Let $(a|b)$ and $(c|d)$ be two conditionals. If $(a|b) \leftrightarrow (c|d)$ then there exists conditionals $(\alpha|\beta)$, $(\chi|\delta)$ and $(e|f)$ such that

$$(\alpha|\beta) \wedge (\chi|\delta) = (0 \mid \beta \vee \delta), \qquad (2.19)$$
$$(\alpha|\beta) \wedge (e|f) = (0 \mid \beta \vee f), \qquad (2.20)$$
$$(\chi|\delta) \wedge (e|f) = (0 \mid \delta \vee f), \qquad (2.21)$$

and

$$(a|b) = (\alpha|\beta) \vee (e|f) \qquad (2.22)$$
$$(c|d) = (\chi|\delta) \vee (e|f) \qquad (2.23)$$

Using the conditional operations and the definition of equivalence of conditionals, equation (2.19) is equivalent to

$$\alpha\beta\delta' \vee \beta'\chi\delta \vee \alpha\beta\chi\delta = 0,$$

which simplifies by Boolean operations and a de Morgan formula to two inequalities:

$$\alpha\beta(\delta' \vee \chi\delta) \vee (\beta' \vee \alpha\beta)\chi\delta = 0,$$
$$\alpha\beta(\delta' \vee \chi\delta) = 0 = (\beta' \vee \alpha\beta)\chi\delta,$$
$$\alpha\beta \leq \chi'\delta \text{ and } \chi\delta \leq \alpha'\beta. \qquad (2.24)$$

Similarly, equations (2.20) and (2.21) are equivalent to

$$\alpha\beta \leq e'f \text{ and } ef \leq \alpha'\beta, \qquad (2.25)$$

$$\chi\delta \le e'f \text{ and } ef \le \chi'\delta. \qquad (2.26)$$

Equations (2.22) and (2.23) are respectively equivalent to the following two lines:

$$ab = \alpha\beta \vee ef \text{ and } b = \beta \vee f, \qquad (2.27)$$
$$cd = \chi\delta \vee ef \text{ and } d = \delta \vee f \qquad (2.28)$$

Starting with (2.27),

$$\begin{aligned} b = \beta \vee f &= \alpha\beta \vee \alpha'\beta \vee ef \vee e'f \\ &= ab \vee \alpha'\beta \vee e'f \\ &\ge ab \vee \chi\delta \\ &\ge \chi\delta \end{aligned}$$

using first (2.27) and half of both (2.24) and (2.26).

Since also $b = \beta \vee f \ge f \ge ef$, therefore $b \ge \chi\delta \vee ef = cd$. That is, $cd \le b$. By symmetry, $ab \le d$. Therefore both $ab \le d$ and $cd \le b$. That proves the first half of Theorem 3.2.

Conversely, suppose (a|b) and (c|d) are two conditionals such that $ab \le d$ and $cd \le b$. Let

$$\begin{aligned} (\alpha|\beta) &= (ac'|b), \\ (\chi|\delta) &= (a'c|d), \\ (e|f) &= (ac|bd). \end{aligned}$$

Then

$$\begin{aligned} (\alpha|\beta) \wedge (e|f) &= (ac'|b) \wedge (ac|bd) \\ &= ((ac')(b)(bd)' \vee (b')(ac)(bd) \vee (ac')(b)(ac)(bd) \mid b \vee bd) \\ &= (ac'bd' \vee 0 \vee 0 \mid b) = (0|b) \end{aligned}$$

using that $ab \le d$, that is, $abd' = 0$. So $(\alpha|\beta)$ and (e|f) are orthogonal. By symmetry, so too are $(\chi|\delta)$ and (e|f). Also $(\alpha|\beta)$ and $(\chi|\delta)$ are orthogonal because $(\alpha|\beta) \wedge (\chi|\delta) = (ac'|b) \wedge (a'c|d) = (ac'bd' \vee b'a'cd \vee ac'ba'cd \mid b \vee d) = (0 \mid b \vee d)$, using that both $abd'$ and $b'cd = 0$.

Finally, $(\alpha|\beta) \vee (e|f) = (ac'|b) \vee (ac|bd) = (ac'b \vee acbd \mid b \vee bd) = (ac'b \vee abc \mid b) = (a|b)$, using that $ab \le d$ in the second to last step. So $(a|b) = (\alpha|\beta) \vee (e|f)$.

By symmetry, $(c|d) = (\chi|\delta) \vee (e|f)$. Therefore $(a|b) \leftrightarrow (c|d)$. That ends the proof of Theorem 3.2.

**3.3 Corollary on Simultaneous Verifiability.** $(a|b) \leftrightarrow (c|d)$ iff $(a|b) \wedge (c|d) = (abcd \mid b \vee d)$.

The corollary follows easily by computing $(a|b) \wedge (c|d)$ using the conjunction operation and noting that the terms $abd'$ and $cdb'$ both are 0 leaving the term $abcd$ alone in the consequent.

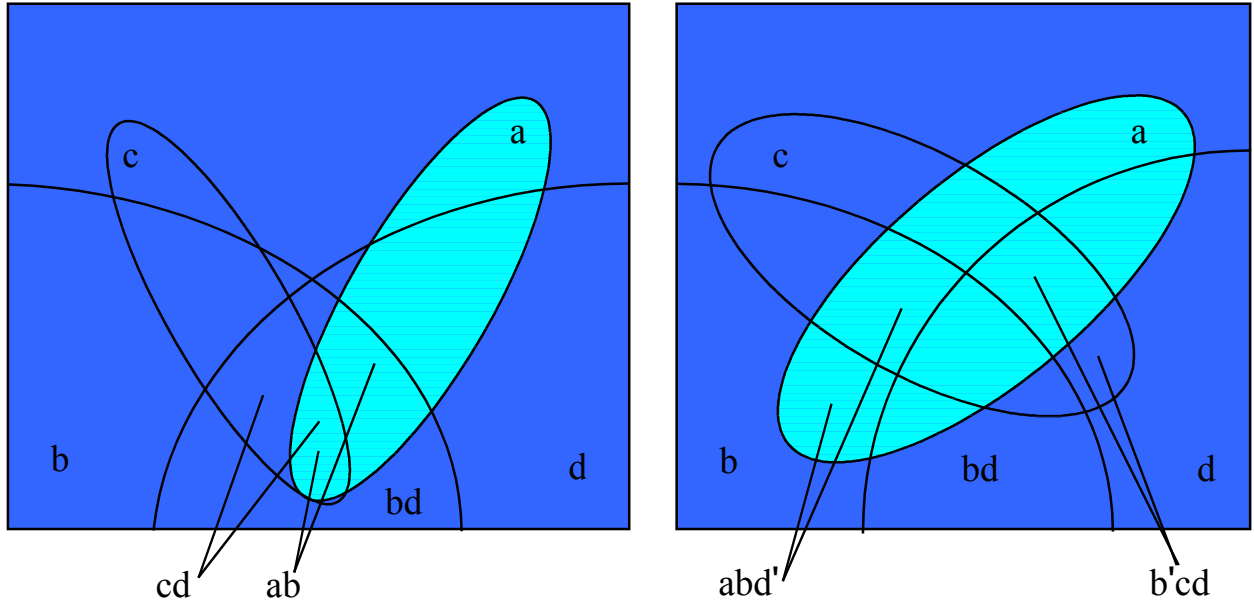
**Figure 2.** Left: Simultaneous Verifiability for Conditional Events (a|b) & (c|d)

If (a|b) and (c|d) are *simultaneously verifiable* then according to the theorem the truth of either conditional implies the applicability of the other conditional, and so if either is true then the other can be verified as either true or false. However if one conditional is false, nothing can be said about the other conditional. The following definition remedies this situation.

**3.4 Definition (Simultaneous Falsifiability).** Two conditionals (a|b) & (c|d) are *simultaneously falsifiable* if and only if their negations, (a'|b) & (c'|d), are simultaneously verifiable. That is, (a'|b) $\leftrightarrow$ (c'|d).

**3.5 Corollary on Simultaneous Falsifiability.** In the conditional event logic two conditionals (a|b) & (c|d) are simultaneously falsifiable if and only if a'b $\leq$ d and c'd $\leq$ b, that is, is the falsity of one conditional implies the applicability of the other conditional.

**3.6 Corollary on Simultaneous Verifiability and Falsifiability.** Two conditionals (a|b) & (c|d) are simultaneously verifiable and simultaneously falsifiable if and only if b = d.

**Proof of Corollary 3.6.** By the Theorem ab $\leq$ d and cd $\leq$ b, and also a'b $\leq$ d and c'd $\leq$ b. So obviously b = ab $\vee$ a'b $\leq$ d and similarly d = cd $\vee$ c'd $\leq$ b. So b=d.

In various ways many authors have expressed the fact when quantum variables exist in a common Boolean algebra they are completely classical. For conditional event algebra, that is, for the algebra of Boolean fractions, this corresponds to the two variables having equal conditions.

**3.7 Theorem on Boolean Sub-Algebras.** Two conditionals (a|b) and (c|d) are in a common Boolean sub-algebra of the (non-Boolean) algebra of conditionals if and only if b = d $\neq$ 0.

**Proof of Theorem 3.7.** Suppose (a|b) & (c|d) are in a common Boolean algebra $\mathcal{A}$. Then neither b nor d can be 0 since otherwise U = (1|0) would be in $\mathcal{A}$. But if U ∈ $\mathcal{A}$, then U has a complement (negation) X such that X ∨ U = 1 and X ∧ U = 0. But from the operations on conditionals, or from the 3-valued truth tables of Table 1, X ∨ U = X and X ∧ U = X. Therefore 1 = X ∨ U = X = X ∧ U = 0, which by definition is impossible in a Boolean algebra because 1 ≠ 0.

Since b ≠ 0, then suppose ω is an instance of b taken as a Boolean event (or as an atom in the logical terminology). That is, suppose ω ∈ b. Now either ω ∈ a or ω ∈ a'. Let ω ∈ a by renaming a' and a, if necessary, because if (a|b) ∈ $\mathcal{A}$ then its negation (a'|b) is also in $\mathcal{A}$, and so the proof could have started with either of them. Next consider the expression

$$(a|b) \wedge [(a'|b) \vee (c|d)].$$

For ω it has truth-value T ∧ [F ∨ t(c|d)], where t(c|d) denotes the truth-value of (c|d) for ω. By cases it follows easily that

$$T \wedge [F \vee t(c|d)] = \begin{cases} F, \text{ if } t(c \mid d) = F \text{ or } U, \\ T, \text{ if } t(c \mid d) = T. \end{cases}$$

On the other hand, [(a|b) ∧ (a'|b)] ∨ [(a|b) ∧ (c|d)] has truth-value

$$[T \wedge F] \vee [T \wedge t(c|d)] = \begin{cases} F, \text{ if } t(c \mid d) = F, \\ T, \text{ if } t(c \mid d) = T \text{ or } U. \end{cases}$$

But these two expressions must be equal because $\mathcal{A}$ is a (distributive) Boolean algebra. So t(c|d) ≠ U. So ω ∈ d. Thus any ω in b is also in d. That is, b ≤ d. By symmetry, d ≤ b. Thus b=d.

Conversely, if b=d, then (a|b) and (c|d) are in the common Boolean algebra ($\mathcal{B}$|b) = {(x|b): x ∈ $\mathcal{B}$} since the operations on conditionals all reduce to Boolean operations when the antecedents are equal, and (1|b) and (0|b) are the unity and zero elements of ($\mathcal{B}$|b). That completes the proof of the theorem.

**3.8 Corollary on Boolean Sub-Algebras.** Two conditionals (a|b) & (c|d) are simultaneously verifiable and falsifiable if and only if they are in a common Boolean sub-algebra.

If (a|b) and (c|d) are two conditionals and b ≤ d, then (c|d) is verifiably true or false whenever (a|b) is true or false. That is, if (a|b) is applicable then (c|d) is applicable. This has been denoted (a|b) $\leq_{ap}$ (c|d) in [Cal02] as part of a hierarchy of eleven deductive relations on conditionals.

**3.9 Theorem on Uniqueness of Relative Negation.** Let (a|b) and (c|d) be two conditionals. Then (a|b) ∧ (c|d) = (0 | b ∨ d) and (a|b) ∨ (c|d) = (1 | b ∨ d) if and only if b=d and (c|d) = (a|b)'.

**Proof of theorem 3.9.** Using the operations, the first equation is equivalent to abd' ∨ b'cd ∨ abcd = 0, and the second equation is equivalent to ab ∨ cd = b ∨ d. So abd' = 0, b'cd = 0, and abcd = 0.

Conjunction on both sides of the equation (ab ∨ cd = b ∨ d) by d′ yields 0 = bd′. So b ≤ d. By symmetry d ≤ b. Therefore b=d. So ab ∨ cb = b. But (ab ∨ cb = b) ⇔ (b ≤ a ∨ c) ⇔ (ba′c′ = 0) ⇔ a′b ≤ c. Also (abc = 0) ⇔ (c ≤ (ab)′) ⇔ (c ≤ a′b ∨ b′). Therefore, a′b ≤ c ≤ a′b ∨ b′. Since c is between a′b and (a′b ∨ b′), c = a′b ∨ xb′ for some proposition x. So (c|d) = (a′b ∨ xb′ | b) = (a′b ∨ xb′b |b) = (a′|b) = (a|b)′. Conversely, if b=d and (c|d) = (a|b)′, then clearly (a|b) ∧ (c|d) = (a|b) ∧ (a|b)′ = (0|b) = (0 | b ∨ d) and (a|b) ∨ (c|d) = (a|b) ∨ (a|b) ′ = (1|b) = (1 | b ∨ d). That completes the proof of theorem 3.9.

**3.10 Hilbert Space and Orthoalgebras.** Indicating a still rich field of research Coecke et al [Coe01, p17] listed two paragraphs of recent efforts by various authors to generalize orthoalgebras and lattices.

They defined (p16) an orthoalgebra as a pair (*L*, ⊕) where *L* is a set and ⊕ is a commutative, associative, partial operation on *L* such that

(1)   There exists 0 ∈ *L* such that ∀p ∈ *L*, p ⊕ 0 = p,
(2)   There exists 1 ∈ *L* such that ∀p ∈ *L*, there exists a unique q ∈ *L* such that p ⊕ q = 1,
(3)   p ⊕ p = 0 if it exists.

In the algebra of conditionals there is a zero for each condition, which makes the above framework come out a little differently.

**3.11 Theorem on Orthoalgebras.** If (a|b) ⊕ (c|d) is defined to be the conditional event (abc′d ∨ a′bcd | b ∨ d) then

(1)   (a|b) ⊕ (0|b) = (a|b),
(2)   (a|b) ⊕ (a′|b) = (1|b), and (a′|b) is unique, and
(3)   (a|b) ⊕ (a|b) = (0|b).

**Proof of Theorem 3.11.** Except for uniqueness in (2), items (1)–(3) of the Theorem are obvious from the definition of ⊕. Suppose that (c|d) is another complement of (a|b) satisfying (2). That is, suppose (a|b) ⊕ (c|d) = (1|b). Then (abc′d ∨ a′bcd | b ∨ d)  = (1|b). So by the definition of equal conditionals, b ∨ d = b and abc′d ∨ a′bcd = b. Since b ∨ d = b, therefore d ≤ b. Now abc′d ∨ a′bcd = b is equivalent to (ac′ ∨ ca′)bd = b, which means b ≤ (ac′ ∨ ca′)d. This is equivalent to b[(ac′ ∨ ca′)d]′ = 0. That is, 0 = b[(ac′ ∨ ca′)′ ∨ d′] = b[(a′ ∨ c)(c′ ∨ a) ∨ d′] = b[a′c′ ∨ ac ∨ d′] = a′bc′ ∨ abc ∨ bd′. Therefore a′bc′ = 0, abc = 0, and bd′ = 0. But bd′ = 0 means b ≤ d. So b=d. Furthermore, (a′bc′ = 0) is equivalent to (a′b ≤ c), and (abc = 0) is equivalent to c ≤ (ab)′ = (a′ ∨ b′) = (a′b ∨ b′). Combining these results about c, we have a′b ≤ c ≤ (a′b ∨ b′). But then (c|d) = (a′b ∨ xb′ | b) for some x ∈ ℬ. So (c|d) = (a′b ∨ xb′ | b) = (a′b ∨ xb′b | b) = (a′b|b) = (a′|b). Therefore uniqueness is proved, and that completes the proof of the theorem.

**3.12 Orthocomplementations.** [Pir76, 8]: Piron following Varadarajan [Var68] defined a lattice as orthocomplemented if there is a mapping that assigns to each element b of the lattice a negation element b′ such that for any elements b and c:

1) $(b')' = b$,
2) $b \wedge b' = 0$ and $b \vee b' = 1$,
3) $b \leq c$ implies $c' \leq b'$.

This definition is expanded (p30) to a "relative" orthocomplementation by restricting all elements to some fixed element d so that

1) $(b')'d = bd$,
2) $bd \wedge b'd = 0$ and $bd \vee b'd = d$,
3) $bd \leq cd$ implies $c'd \leq b'd$.

It seems clear that this concept is very close to the relative complementation in ($\mathcal{B}|\mathcal{B}$).

**3.13 Projections and the Sasaki Mapping.** Coecke [Coe02, p3] in covering Piron's work defined the two so-called Sasaki projections [Sas55] that represent the change from an initial state "a" to either of the states

$$\phi_b(a) = b \wedge (b' \vee a) \quad \text{or} \quad \phi_{b'}(a) = b' \wedge (b \vee a),$$

where b and b' are "eigenproperties". These are projections of a onto either b or b'.

Working in an orthomodular lattice Delmas-Rigoutsos [Del97, p59] also defined the first of these projections as a non-commutative operation (∘) as follows:

$$b \circ a = b \wedge (b' \vee a).$$

Piron [Pir76, 69] defined this Sasaki projection mapping on the atomic propositions and he showed it has the following properties for all propositions a, b and c:

1) $\phi_b(a) = a \Leftrightarrow a \leftrightarrow b$,
2) $\phi_b(a) = 0 \Leftrightarrow a \leftrightarrow b'$,
3) $\phi_b(\phi_b(a)) = \phi_b(a)$,
4) $\phi_c(\phi_b(a)) = \phi_{b \wedge c}(a) \Leftrightarrow c \leftrightarrow b$,
5) $\phi_c(\phi_b(a)) = \phi_b(\phi_c(a)) \Leftrightarrow c \leftrightarrow b$.

Interpreting the propositions above as conditionals instead of orthomodular lattice propositions, and adopting the simpler notation of Delmas-Rigoutsos [Del97], we have the following definition.

**3.14 Definition of Sasaki Projections for Conditionals.** If $(a_1|a_2)$ and $(b_1|b_2)$ are two conditionals then the Sasaki projection of $(a_1|a_2)$ onto $(b_1|b_2)$ is

$$(b_1|b_2) \circ (a_1|a_2) = (b_1|b_2) \wedge [(b_1|b_2)' \vee (a_1|a_2)].$$

Combining the right hand side into a single conditional with the operations on conditionals yields

$$(b_1|b_2) \circ (a_1|a_2) \;=\; (a_1 a_2(b_2' \vee b_1) \mid a_2 \vee b_2).$$

Since $(\boldsymbol{\mathcal{B}}|\boldsymbol{\mathcal{B}})$ is not always distributive, $\circ$ defines a non-commutative operation on $(\boldsymbol{\mathcal{B}}|\boldsymbol{\mathcal{B}})$. It satisfies the following counterparts of Peron's results above.

**3.15 Theorem on Sasaki Projections.** If $(a_1|a_2)$, $(b_1|b_2)$ and $(c_1|c_2)$ are three conditionals then

i)       $(b_1|b_2) \circ (a_1|a_2) \;=\; (a_1|a_2)$ if and only if $b_2 \leq a_2$ and $b_1'b_2 \leq a_1'a_2$,
ii)      $(b_1|b_2) \circ (a_1|a_2) \;=\; (0 \mid a_2 \vee b_2)$ if and only if $a_1 a_2 \leq b_1'b_2$,
iii)     $(b_1|b_2) \circ [(b_1|b_2) \circ (a_1|a_2)] \;=\; (b_1|b_2) \circ (a_1|a_2)$,
iv)     $(c_1|c_2) \circ [(b_1|b_2) \circ (a_1|a_2)] = [(b_1|b_2) \wedge (c_1|c_2)] \circ (a_1|a_2)$,
v)      $(c_1|c_2) \circ [(b_1|b_2) \circ (a_1|a_2)] = (b_1|b_2) \circ [(c_1|c_2) \circ (a_1|a_2)]$.

**Proof of Theorem 3.15.** Concerning i), $(b_1|b_2) \circ (a_1|a_2) \;=\; (a_1|a_2)$ if and only if $(a_1 a_2(b_2' \vee b_1) \mid a_2 \vee b_2) = (a_1|a_2)$. Equivalently, $a_2 \vee b_2 = a_2$ and $a_1 a_2(b_2' \vee b_1) = a_1 a_2$. That is, $b_2 \leq a_2$ and $a_1 a_2 \leq b_2' \vee b_1$. Taking complements on both sides of the second of these inequalities yields the equivalent inequality $b_1'b_2 \leq a_1'a_2 \vee a_2'$. Since $b_2 \leq a_2$, conjunction on both sides of $b_1'b_2 \leq a_1'a_2 \vee a_2'$ by $a_2$ yields $b_1'b_2 \leq a_1'a_2$. Conversely, the latter inequality implies $b_1'b_2 \leq a_1'a_2 \vee a_2'$. So the equivalence is maintained. [Item i) was already proved earlier in a different form as the second corollary to the modular law.] Concerning ii), $(b_1|b_2) \circ (a_1|a_2) \;=\; (0 \mid a_2 \vee b_2)$ if and only if $a_1 a_2(b_2' \vee b_1) = 0$, which is equivalent to $a_1 a_2 \leq b_1'b_2$. Item iii) of the theorem follows immediately from the simplification of $(b_1|b_2) \circ (a_1|a_2)$ to $(a_1 a_2(b_2' \vee b_1) \mid a_2 \vee b_2)$ since $(b_1|b_2) \circ (a_1 a_2(b_2' \vee b_1) \mid a_2 \vee b_2) = (a_1 a_2(b_2' \vee b_1)(b_2' \vee b_1) \mid a_2 \vee b_2 \vee b_2) = (b_1|b_2) \circ (a_1|a_2)$. Concerning iv), $[(b_1|b_2) \wedge (c_1|c_2)] \circ (a_1|a_2) = [(b_2' \vee b_1)(c_2' \vee c_1) \mid b_2 \vee c_2)] \circ (a_1|a_2) = \{a_1 a_2(b_2' \vee b_1)(c_2' \vee c_1)[(b_2 \vee c_2)(c_2' \vee c_1) \vee b_2'c_2'] \mid a_2 \vee b_2 \vee c_2\} = \{a_1 a_2(b_2' \vee b_1)(c_2' \vee c_1) \mid a_2 \vee b_2 \vee c_2\} = (c_1|c_2) \circ [(b_1|b_2) \circ (a_1|a_2)]$. Concerning v), since conjunction ($\wedge$) is commutative on conditionals item iv) easily implies item v). That completes the proof of the theorem.

**3.16 Corollary on Sasaki Projections.** If $(a_1|a_2)$ and $(b_1|b_2)$ are two conditionals then

i)       $(b_1|b_2) \circ (a_1|a_2) \;=\; (a_1|a_2)$       $\Leftrightarrow$      $(a_1|a_2) \leq_\wedge (b_1|b_2)$,
ii)      $(b_1|b_2) \circ (a_1|a_2) \;=\; (0 \mid a_2 \vee b_2)$    $\Leftrightarrow$      $(a_1|a_2) \leq_{tr} (b_1|b_2)'$,
iii)     $(b_1|b_2) \circ (b_1|b_2) = (b_1|b_2)$; the operation $\circ$ is an idempotent operation,

**Proof of Corollary 3.16.** In [Cal91, pp86-87] the deductive relation $(a_1|a_2) \leq_\wedge (b_1|b_2)$ was defined by the equation $(a_1|a_2) \wedge (b_1|b_2) = (a_1|a_2)$, and it was found to be equivalent to $a_2' \leq b_2'$ and $a_1 \vee a_2' \leq b_1 \vee b_2'$. But, these are equivalent to $b_2 \leq a_2$ and $b_1'b_2 \leq a_1'a_2$. So i) of the corollary follows from i) of the theorem. Concerning ii), the deductive relation $(a_1|a_2) \leq_{tr} (b_1|b_2)$ was defined in [Cal96] and [Cal02] by the inequality $a_1 a_2 \leq b_1 b_2$; the truth of $(a_1|a_2)$ implies the truth of $(b_1|b_2)$. So item ii) of the corollary follows from ii) of the theorem. Item iii) follows directly from the definition of $(b_1|b_2) \circ (a_1|a_2)$. That completes the proof of the corollary.

Notice that unlike Peron's corresponding results the equations of iv) and v) above are always satisfied without any additional conditions. Therefore, as a way of distinguishing compatible pairs of conditionals these equations appear to be inadequate. Or an alternate definition of $(b_1|b_2) \circ (a_1|a_2)$ may be needed.

However, referring to the relationship of iii) below as the *compatibility* relation, denoted here by $a \asymp b$, Delmas-Rigoutsos [Del97, pp59-60] proved the following results in an orthomodular lattice:

i)      $b \vee a = b \vee (b' \circ a)$,
ii)     $c \circ (a \vee b) = (c \circ a) \vee (c \circ b)$,
iii)     $b \circ a = a \circ b \quad \Leftrightarrow \quad b \circ a = b \wedge a \quad \Leftrightarrow \quad b \circ a \leq a$.
iv)     $b \asymp a \quad \Leftrightarrow \quad b' \asymp a$,
v)      If $b \asymp a_i$ for each i, then $b \asymp (\vee_i a_i)$ and $b \asymp (\wedge_i a_i)$.

As indicated below, for conditionals the relationship $b \circ a = a \circ b$ is equivalent to simultaneous verifiability ($\leftrightarrow$) as defined earlier. So two conditionals, $(a|b)$ & $(c|d)$, will be called *compatible* ($\asymp$) in case both $(a|b) \leftrightarrow (c|d)$ and $(a|b)' \leftrightarrow (c|d)'$. This was shown in Corollary 3.6 to be equivalent to $b=d$.

The deductive relation $\leq_{bo}$ was defined in [Cal96] and [Cal02] by $(a|b) \leq_{bo} (c|d)$ if and only if $b=d$ and $ab \leq cd$. As such, $\leq_{bo}$ is just Boolean deduction restricted to conditionals with equivalent antecedents. It is Boolean deduction in the Boolean algebra $(\mathcal{B}|b) = \{(a|b): a \in \mathcal{B}\}$. Exploring these relationships in the context of conditionals there are the following parallel results.

**3.17 Theorem on Compatibility.** If $(a_1|a_2)$, $(b_1|b_2)$ and $(c_1|c_2)$ are three conditionals, and if $\{(a_i|b_i)\}$ is a countable, indexed set of conditionals, then

i)      $(b_1|b_2) \vee (a_1|a_2) = (b_1|b_2) \vee [(b_1|b_2)' \circ (a_1|a_2)]$,
ii)     $(c_1|c_2) \circ [(b_1|b_2) \vee (a_1|a_2)] = (c_1|c_2) \circ (b_1|b_2) \vee (c_1|c_2) \circ (a_1|a_2)$,
iii)
       a.    $(b_1|b_2) \circ (a_1|a_2) = (a_1|a_2) \circ (b_1|b_2) \quad \Leftrightarrow \quad (b_1|b_2) \leftrightarrow (a_1|a_2)$,
       b.    $(b_1|b_2) \circ (a_1|a_2) = (b_1|b_2) \wedge (a_1|a_2) \quad \Leftrightarrow \quad b_1 b_2 \leq a_2$,
       c.    $(b_1|b_2) \circ (a_1|a_2) \leq_{bo} (a_1|a_2) \quad \Leftrightarrow \quad b_2 \leq a_2$,
iv)     $[(b_1|b_2) \asymp (a_1|a_2) \Leftrightarrow (b_1|b_2)' \asymp (a_1|a_2)] \quad \Leftrightarrow \quad b_2 \leq a_2$,
v)      If $(c|d) \asymp (a_i|b_i)$ for each i, then $(c|d) \asymp \vee_i (a_i|b_i)$ and $(c|d) \asymp \wedge_i (a_i|b_i)$.

**Proof of Theorem 3.17.** Concerning i), $[(b_1|b_2)' \circ (a_1|a_2)] = (a_1 a_2 (b_2' \vee b_1') \mid a_2 \vee b_2) = (a_1 a_2 (b_2 b_1)' \mid a_2 \vee b_2)$. So $(b_1|b_2) \vee [(b_1|b_2)' \circ (a_1|a_2)] = (b_1 b_2 \vee (a_1 a_2)(b_2 b_1)' \mid a_2 \vee b_2) = (b_1 b_2 \vee a_1 a_2 \mid a_2 \vee b_2) = (b_1|b_2) \vee (a_1|a_2)$. Concerning ii), $(c_1|c_2) \circ [(b_1|b_2) \vee (a_1|a_2)] = (c_1|c_2) \circ (b_1 b_2 \vee a_1 a_2 \mid b_2 \vee a_2) = (b_1 b_2 \vee a_1 a_2)(c_1 \vee c_2') \mid c_2 \vee b_2 \vee a_2) = [(b_1 b_2)(c_1 \vee c_2') \vee (a_1 a_2)(c_1 \vee c_2') \mid c_2 \vee b_2 \vee c_2 \vee a_2] = [(b_1 b_2)(c_1 \vee c_2') \mid c_2 \vee b_2] \vee [(a_1 a_2)(c_1 \vee c_2') \mid c_2 \vee a_2] = (c_1|c_2) \circ (b_1|b_2) \vee (c_1|c_2) \circ (a_1|a_2)$. Concerning item iii)

part a, $(b_1|b_2) \circ (a_1|a_2) = (a_1|a_2) \circ (b_1|b_2)$ if and only if $a_1a_2(b_2' \vee b_1) = b_1b_2(a_2' \vee a_1)$. Conjoining both sides by $a_2'$ yields $b_1b_2a_2' = 0$, that is, $b_1b_2 \leq a_2$. Similarly, or by symmetry, $a_1a_2 \leq b_2$. Conversely, if $b_1b_2a_2' = 0 = a_1a_2b_2'$, then $(b_1|b_2) \circ (a_1|a_2) = (a_1a_2b_1b_2 \mid a_2 \vee b_2) = (a_1|a_2) \circ (b_1|b_2)$, which proves iii) part a. Concerning item iii) part b, $(b_1|b_2) \circ (a_1|a_2) = (b_1|b_2) \wedge (a_1|a_2)$ if and only if $a_1a_2(b_2' \vee b_1) = b_1b_2a_2' \vee a_1a_2b_2' \vee b_1b_2a_1a_2$. Conjunction on both sides by $a_2'$ yields $b_1b_2a_2' = 0$, which is equivalent to $b_1b_2 \leq a_2$, and conversely the latter formula easily implies iii) part b. Concerning iii) part c, $(b_1|b_2) \circ (a_1|a_2) \leq_{bo} (a_1|a_2) \Leftrightarrow (a_1a_2(b_2' \vee b_1) \mid a_2 \vee b_2) \leq_{bo} (a_1|a_2) \Leftrightarrow (a_2 \vee b_2) = a_2$ and $a_1a_2(b_2' \vee b_1) \leq a_1a_2 \Leftrightarrow b_2 \leq a_2$. Concerning item iv), $(b_1|b_2) \asymp (a_1|a_2)$ just in case both $a_1a_2 \leq b_2$ and $b_1b_2 \leq a_2$. Similarly, $(b_1|b_2)' \asymp (a_1|a_2)$ just in case $b_1'b_2 \leq a_2$ and $a_1a_2 \leq b_2$. So if $(b_1|b_2) \asymp (a_1|a_2)$ holds then $(b_1|b_2)' \asymp (a_1|a_2)$ holds only if $b_1'b_2 \leq a_2$ in addition to $b_1b_2 \leq a_2$, in which case $b_2 = b_1b_2 \vee b_1'b_2 \leq a_2$. Conversely, if $b_2 \leq a_2$, then both $b_1b_2 \leq a_2$ and $b_1'b_2 \leq a_2$. So if in addition $a_1a_2 \leq b_2$ then both $(b_1|b_2) \asymp (a_1|a_2)$ and $(b_1|b_2)' \asymp (a_1|a_2)$. Concerning v), if $(c|d) \asymp (a_i|b_i)$ for each i, then for each i, $cd \leq b_i$ and $a_ib_i \leq d$. So $cd \leq (\vee_i b_i)$ and $(\vee_i a_ib_i) \leq d$. So $(c|d) \asymp (\vee_i a_ib_i \mid \vee_i b_i) = \vee_i (a_i|b_i)$. That shows the first half of v). Since the reduced consequent, $[\wedge_i (a_i \vee b_i')] \wedge (\vee_j b_j)$ of a conjunction $\wedge_i (a_i|b_i)$ is always below ($\leq$) the reduced consequent $\vee_i a_ib_i$ of the corresponding disjunction $\vee_i (a_i|b_i)$, therefore $[\wedge_i (a_i \vee b_i')] \wedge (\vee_j b_j) \leq (\vee_i a_ib_i) \leq d$ and $cd \leq (\vee_i b_i)$. So $(c|d) \asymp \wedge_i (a_i|b_i)$. That completes the proof of the theorem.

**3.18. Distinctions Possible with Boolean Fractions.** It is significant that conditionals (Boolean fractions) can often distinguish relationships that are not distinguishable in the orthodox quantum formulations. For instance, as shown in iv) of the last theorem, or directly, $(b_1|b_2)' \asymp (a_1|a_2)$ is equivalent to $b_1'b_2 \leq a_2$ and $a_1a_2 \leq b_2$, which means that the truth of $(a_1|a_2)$ implies the applicability of $(b_1|b_2)$ and the falsity of $(b_1|b_2)$ implies the applicability of $(a_1|a_2)$. Thus one might wish to discuss several kinds of simultaneous verifiability for two conditionals $(a_1|a_2)$ and $(b_1|b_2)$ such as:

1. $a_1a_2 \leq b_2$; if $(a_1|a_2)$ is true then $(b_1|b_2)$ is applicable (and so verifiable).
2. $a_1'a_2 \leq b_2$; if $(a_1|a_2)$ is false then $(b_1|b_2)$ is applicable (and so verifiable).
3. $a_1a_2 \leq b_2$ and $b_1b_2 \leq a_2$; if either conditional is true then the other is applicable. In this paper this has been called *simultaneous verifiability* and denoted by $(a_1|a_2) \leftrightarrow (b_1|b_2)$.
4. $b_1'b_2 \leq a_2$ and $a_1'a_2 \leq b_2$; if either conditional is false then the other is applicable. In this paper this has been called *simultaneous falsifiability*: $(a_1|a_2)' \leftrightarrow (b_1|b_2)'$.
5. $a_1'a_2 \leq b_2$ and $b_1b_2 \leq a_2$; if the first conditional is false then the second conditional is applicable, and if the second conditional is true then the first is applicable. This combines the relationships in 1. and 2., reversing the notation in 2. This is $(a_1|a_2)' \leftrightarrow (b_1|b_2)$.
6. $a_2 \leq b_2$; if $(a_1|a_2)$ is applicable then $(b_1|b_2)$ is applicable (and so verifiable). This is 1. and 2.
7. $a_2 = b_2$; equivalent conditions. This has been called *compatibility*. Note $(a_1|a_2)' \leftrightarrow (b_1|b_2)$ and $(a_1|a_2) \leftrightarrow (b_1|b_2)'$ both hold if and only if b=d. Also 3. and 4. hold just in case $a_2 = b_2$.

**3.19 True, Wholly True and Given True.** One of the main problems confronting anyone trying to reformulate quantum mechanics is how the truth of a quantum proposition is defined as compared to in Boolean logic or probability logic. While a Boolean proposition b is supposed to be either true or false, with nothing in between, that only applies to each instance, occurrence or atomic state ω. Usually in general probability theory a moving occurrence ω determines whether a fixed proposition "occurs" or is "true" depending on whether ω ∈ b or ω ∈ b′. The propositions or events b are temporarily fixed while the instances ω are changing, being any one of a sample space of outcomes. Thus a proposition may be true in one instance and false in another. In the indicator representation of conditionals the truth of b for ω can be written as b(ω) = 1.

Now if b is true for all occurrences ω in the sample space Ω then this can be expressed as "b≡1", or simply b=1. This is the situation when b is a certainty, when b is wholly true.

Still another type of truth for b is "given b", which is a restriction of the set Ω to those instances for which b is true. This can be expressed as (x | b) or (· | b), "given b", meaning any member of the Boolean algebra (𝓑|b) of events x each conjoined with b.

While the probability P(b) of b can be one number, the P(b=1) that b is wholly true can be another number [Cal87, p221][Cal94, p1683], and of course the conditional probability P(· | b) when b is "given" is something else again.

In quantum logic, a pure (atomic) state ω may be considered fixed, so that the truth of a moving proposition b depends upon whether b includes ω. The events can move while the state ω is fixed. Thus, given a pure state ω a quantum proposition b is either wholly true or wholly false depending on whether it happens to include the fixed state ω.

Actually, in practice both the propositions and the states move, especially in quantum mechanics, where determining the truth of a proposition b generally changes the state as a result of the measurement.

**3.20 Quantum Operations and Truth Revisited.** As mentioned earlier, G. Schay [Sch68] observed that the disjunction and conjunction operations on conditionals promoted here seem to correspond to the usual meaning of "or" and "and", but he mostly passes them by for operations that form a lattice. Using the conjunction operation, for instance, to define a partial order in the usual way (that is, b ≤ c means b ∧ c = b) Schay undoubtedly discovered that (a|b) ∧ (c|d) is the GLB of its components but (a|b) ∨ (c|d) is not the LUB. So he chose a different disjunction operation to go with ∧. Similarly, he chose a different conjunction operation to go with ∨.

However, as Schay himself showed, the resulting lattices are distributive, making them inadequate for representing quantum operations. While the properties of a lattice are not essential for representing quantum operations, the property of non-distributivity is.

In their account of the quantum operations Aerts et al [Aer00A] carefully explained how a quantum state p represented by a unit vector $v_p$ determines the truth of a quantum proposition b as it is contained (or not) in the closed linear subspace $M_b$ representing b in the Hilbert space. They pointed out that although conjunction and implication behave as they do in the classical case, disjunction of two propositions, a and b, does not. While a ∧ b corresponds to $M_a \cap M_b$ and a ≤ b (a implies b) corresponds to $M_a \subset M_b$, the disjunction a ∨ b corresponds to $cl(M_a \cup M_b)$, the *closure* of the union of the closed subspaces $M_a$ and $M_b$. This closed subspace is the LUB of $M_a$ and $M_b$. It is likely that Schay [Sch68] was influenced by this LUB connection when he chose his operations on conditionals.

However it is possible to recreate similar relationships in the algebra of conditionals while maintaining the intuitive operations promoted in this paper. For instance, while the intersection of two (sum) ideals, I and J, of propositions is also a sum ideal, the union is generally not an ideal. The LUB of two sum ideals is generally larger than the simple union of the individual ideals [Cal87, p208].

This sum ideal concept was expanded in [Cal96, p207] and especially in [Cal02] to include deductively closed sets of conditional events with respect to some naturally arising deductive relations on conditionals. If $\leq_x$ is a deductive relation (reflexive and transitive), also called a preorder on the conditionals and J is a set of conditionals, then $H_x(J)$ is defined as the smallest set of conditionals that includes J and which is closed under conjunction and deduction with respect to $\leq_x$. This is the LUB of J with respect to $\leq_x$.

A remarkable fact rediscovered in the deductive analysis [Cal02, p173] but already present in the writings of Adams [Ada75] is that in determining the implications of a finite set of conditionals, sometimes there is something else required besides the simple conjunction of the members of J. For instance, in determining the deductive consequences of a set J = {(a|b),(c|d)} of two conditionals with respect to the deductive relation $\leq_{pm}$, the simple conjunction (a|b) ∧ (c|d) is insufficient to deduce all the implications of J. That is, there are conditionals implied with respect to $\leq_{pm}$ by (a|b), which are not implied by (a|b) ∧ (c|d). Therefore, determining the implications of J with respect to $\leq_{pm}$ requires that first all finite conjunctions of members of J be constructed, and then the union of the implications of these conjunctions will embrace all deductive consequences of J with respect to $\leq_{pm}$. So $H_{pm}\{(a|b),(c|d)\} = H_{pm}\{(a|b), (c|d), (a|b)(c|d)\} = H_{pm}(a|b) \cup H_{pm}(c|d) \cup H_{pm}((a|b)(c|d))$. (For details see [Cal02, pp173-5].)

However, unlike the sum ideals and the deductively closed sets of conditionals, the closed quantum subspaces $M_b$ for a proposition b always include the zero vector 0 as well as the negations of all members. They are subspaces not ideals or deductively closed sets. Thus there seems to be a potential identification or correspondence between these subspaces $M_b$ and the conditional spaces (𝓑|b) for fixed b. For two conditional spaces, (𝓑|b) and (𝓑|d), their LUB would be {(𝓑|b), (𝓑|d), (𝓑| b ∨ d)}, which is closed but in general no longer a Boolean algebra.

Recall that $\leq_{pm}$ is the deductive relation for which (a|b) ⊥ (c|d) is true if and only if (a|b) $\leq_{pm}$ (c|d)′. By contrast, with respect to $\leq_\wedge$, the simple conjunction is sufficient for deducing all implications of two or more conditionals. So $H_\wedge\{(a|b), (c|d)\} = H_\wedge((a|b)(c|d))$. See [Cal02, p173] for details.

The introduction of various deductive relations $\leq_x$ opens up the possibility of defining the orthogonality of two conditionals (a|b) and (c|d) by one of the deductive relations that is stronger than $\leq_{pm}$ such as $\leq_{m\wedge}$, which combines the requirements of $\leq_{pm}$ and $\leq_\wedge$. Another extension is to define the LUB and GLB with respect to various deductive relations $\leq_x$ identified in [Cal02, p163].

These were characterized in terms of Boolean relations independent of the operations on conditionals used to motivate and derive them.

A detailed account of these extensions is beyond the scope of this paper. Suffice it to say that early results in this direction indicate that the compatibility and simultaneous verifiability relationships exhibited here are remarkably robust to such changes in the operations. For instance, if $\leq_x$ is any one of six relatively strong deductive relations ($\leq_\wedge$, $\leq_{pm}$, $\leq_\vee$, $\leq_{m\wedge}$, $\leq_{m\vee}$, $\leq_{bo}$) identified in [Cal02] and if the Sasaki project operator with respect to $\leq_x$ is defined by

$$(b_1|b_2) \circ_x (a_1|a_2) = GLB_x\{(b_1|b_2), LUB_x\{(b_1|b_2)', (a_1|a_2)\}\},$$

then the $\circ_x$ operator is still commutative if and only if $a_2 = b_2$.

It seems likely that the disjunction of two quantum propositions is really more like forming their deductive implications than their simple disjunction.

**4. Conclusion.** Although the results given in this paper are individually not very difficult to prove the idea here is to make complicated things simpler, not to prove difficult theorems. In expressing any concept mathematically the best notation is the one that is most efficient and most suggestive of the underlying relationships being described. After all, notation is supposed to facilitate understanding not obfuscate it. The more notation is unnecessarily complicated in mimicking the relationships described, the more it is part of the problem and not part of the solution. There is ample evidence for believing that the standard notation for quantum mechanics suffers from such a malady. The very active investigation by many authors into alternate expressions and interpretations of quantum phenomena demonstrates that theorists are not satisfied with the present descriptions of quantum mechanics no matter how well the quantum experiments statistically agree with the now orthodox formulation.

As nicely described by H. Green [Gre95, p1] quantum (versus continuous) physics was thrust upon M. Planck [Pla00] in 1900 as he sought to harmonize two energy distribution formulas for black body radiation, one for high frequencies of light and one for low frequencies. Today we know that these energy systems exist in a set of discrete states or configurations, which generally shift when some variable of the system is measured. After Heisenberg stated his principle of indeterminacy for observations, it became clear to von Neumann and Birkhoff that an algebraic structure for quantum mechanics would have to be non-distributive. Since Boolean algebra was wholly distributive they had to look elsewhere for such a non-distributive system and settled on normed vector spaces and subspaces.

Boole had not succeeded in adding a worthy division operation to his other 3 operations, and so Kolmogorov was unable to explicitly incorporate conditional events or conditional propositions into his celebrated axiomatization of probability theory. The concept of conditional events (Boolean fractions) to describe conditional statements finally appeared in the 1960's, but by that time the orthodox quantum formulation was well established and spectacularly successful at predicting the statistics of quantum experiments. It was just as spectacularly unsuccessful at illuminating anyone's physical intuition about what was really happening. The strange quantum interference phenomena suggested to Bohr that no such intuitive or "hidden variable" interpretation is possible for quantum mechanics – the so-called Copenhagen interpretation. Since then people have been regularly revisiting the quantum formulation looking for another way to understand quantum dynamics.

Well, if someone is looking for a simpler, non-distributive system with which to describe quantum propositions, they need look no further than to Boolean fractions. This system of ordered pairs of propositions successfully finessed Lewis's triviality results [Lew76] by providing a near-Boolean logical object (a|b) that could also non-trivially carry the conditional probability P(a|b). W. Rödder [Rod00] successfully used the four operations in a very impressive interactive computer program for calculating maximum entropy solutions, and it also recently provided a natural way to extend numerical operations to real functions with variable domains [Cal03]. This system of conditional events, which did not exist for von Neumann and Birkhoff, now appears to adequately and more intuitively describes quantum relationships.

**5. References & Bibliography**